\documentclass[journal,article,submit,moreauthors,pdftex]{aa}

\usepackage{graphicx}
\setcitestyle{author-year}

\usepackage{color, colortbl}
\usepackage{comment}
\usepackage{rotating}
\usepackage{makecell}
\usepackage{array}
\usepackage{float}
\usepackage{placeins}
\usepackage{url}
\usepackage[colorlinks]{hyperref}
\usepackage{nicematrix}
\usepackage{xcolor}
\usepackage{bm}
\usepackage{standalone}
\usepackage{longtable}
\usepackage{txfonts}

\makeatletter
\newcommand\newtag[2]{#1\def\@currentlabel{#1}\label{#2}}
\makeatother

\newcommand{\HeHp}{{{\rm HeH}}^{+}}
\newcommand{\hd}{{\rm HD}}
\newcommand{\DD}{{\rm D_{2}}}
\newcommand{\Dm}{{\rm D}^{-}}
\newcommand{\Dp}{{\rm D}^{+}}
\newcommand{\mD}{{\rm D}}
\newcommand{\ddp}{{\rm D_{2}^{+}}}
\newcommand{\hdp}{{\rm HD^{+}}}
\newcommand{\me}{{\rm e^{-}}}
\newcommand{\mH}{{\rm H}}
\newcommand{\He}{{\rm He}}
\newcommand{\Hep}{{\rm He^{+}}}
\newcommand{\Hepp}{{\rm He^{++}}}
\newcommand{\Hp}{{\rm H}^{+}}
\newcommand{\Hm}{{\rm H}^{-}}
\newcommand{\mHtp}{{\rm H_{2}^{+}}}
\newcommand{\htp}{{\rm H_{3}^{+}}}
\newcommand{\htdp}{{\rm H_2D}^{+}}
\newcommand{\mHt}{{\rm H_{2}}}

\newcommand{\expf}[3]{\exp \left(#1\frac{#2}{#3}\right)}

\def\x{$\times$}
\def\simless{\mathbin{\lower 3pt\hbox
   {$\rlap{\raise 5pt\hbox{$\char'074$}}\mathchar"7218$}}}
\def\simgreat{\mathbin{\lower 3pt\hbox
   {$\rlap{\raise 5pt\hbox{$\char'076$}}\mathchar"7218$}}}

\begin{document}

   \title{Distortion of the CMB spectrum \\ by the first molecules of the Dark Ages}

   \author{Yu. Kulinich
          \and
          B. Novosyadlyj
          }

   \institute{Astronomical Observatory of Ivan Franko National University of Lviv,
              Kyryla i Methodia str., 8, Lviv, 79005, Ukraine\\
              \email{yuriy.kulinich@lnu.edu.ua}
             }

  \abstract
  {The formation of first stars and galaxies at the Cosmic Dawn
  had been preceded by the chain of primordial chemistry reactions
    during Dark Ages which generated first molecules, mostly H$_2$,
    HD, and HeH$^+$, so crucial for first stars to emerge. These
    molecules absorbed and scattered CMB quanta this way distorting
    CMB spectrum.}
  {Estimate how much bound-bound transitions between the rovibrational
    levels of  H$_2$, HD, and HeH$^+$ molecules contribute to
    the distortion of CMB spectrum in the standard $\Lambda$CDM
    cosmology.}
   {We describe the kinetics of the formation of the first molecules
    with system of 166 chemical reactions involving 20 reagents.  The system of differential equations is solved to describe the
    processes of spontaneous and collisional transitions between
    rovibrational levels of H$_2$, HD, and HeH$^+$ molecules.  The
    populations of rovibrational levels and optical thickness of the gas in transition lines between these levels were used to estimate the differential brightness produced by the first molecules on the cosmic microwave background.}
   {It is shown that the signal from the first molecules in the
    standard $\Lambda$CDM cosmology takes the form of an absorption
    profile in the CMB spectrum and originates from the Dark Ages. The
    absorption profile of H$_2$ molecule: features multiple peaks;
    reaches maximum of $\sim 10^{-3}$ Jy/sr within the
    frequency range from $\sim$ 50 GHz to $\sim$ 120 GHz; and
    major contribution in absorption comes from
    redshifts $300>z>200$.
    The absorption profile of HD molecule: features double peaks;
    reaches maximum of $\sim 10^{-5}$ Jy/sr within the frequency
    range from $\sim$ 40 GHz to $\sim$ 70 GHz; and largest part of
    absorption comes from redshifts
    $300>z>30$.
    The absorption profile of HeH$^+$ ion-molecule: has no features
    and reaches maximum of $\sim 10^{-7}$ Jy/sr within the frequency
    range from $\sim$ 200 GHz to $\sim$ 800 GHz; and absorption comes
    mainly from redshifts range $100>z>4$.
  }
   {}


   \maketitle

\section{Introduction}

Approximately 300,000 years (redshift z$\sim$1100) after the Big Bang
the plasma recombined and cosmic microwave background (CMB) photons
had decoupled from baryonic matter, marking the beginning of the Dark
Ages that lasted until the first stars and galaxies come to being at
$z\sim 20 - 30$.  During this time period, the Universe was filled
with: CMB photons, dark matter (DM), dark energy (DE), and a weakly
ionised gas, almost entirely made up of hydrogen ($\approx$92.4~\% of
total number density), helium-4 ($\approx$7.6~\% \citep{Ave15}), along
with trace amounts of deuterium ($\approx$2.45$\cdot 10^{-3}$~\%
\citep{Coo14}), as well as lithium-7 (as predicted by \citep{Coc17} to
amount to $\approx$5.61$\cdot 10^{-8}$~\% while observed \citep{Sbo10}
to be $\approx$1.58$\cdot 10^{-8}$~\% ). Also DM halos were formed
from the initial over-density perturbations at the time, and
primordial chemical reactions synthesized first molecules, with H$_2$
hydrogen molecules as most abundant
\citep{bk978-0-7503-1425-1ch1bib54, Lepp_2002,
  bk978-0-7503-1425-1ch1bib23, bk978-0-7503-1425-1ch1bib61}, deuterium
hydride molecules HD \citep{Lepp_2002,
  Glover2008, bk978-0-7503-1425-1ch1bib38}, helium
hydride ion-molecule HeH$^+$ \citep{Schleicher:2008,
  bk978-0-7503-1425-1ch1bib7}, and lithium hydride molecules LiH
\citep{bk978-0-7503-1425-1ch1bib25, bk978-0-7503-1425-1ch1bib90,
  bk978-0-7503-1425-1ch1bib6, Schleicher:2008,
  bk978-0-7503-1425-1ch1bib90}.

The estimation of fractions of the molecules is determined by the
primordial chemistry model used, namely the set of chemical reactions
and their rates as functions of gas, electron or radiation
temperatures.  The system of primordial chemistry equations can
include as much as 300 reactions for about 30 reagents
\citep{bk978-0-7503-1425-1ch1bib38}. However, besides the completeness
of reaction set, the accuracy of molecular number predictions also
depends on the accuracy of reactions' rates.  Since any updates to the
reaction chains and rates can affect the estimation of abundances,
numerous attempts to trace chemical evolution during the Dark Ages
have yielded quite different results, sometimes by orders of magnitude
\citep{signore2009cosmic,bovino2009fast,Galli1998,galli2002deuterium,
  black2006chemistry,sethi2008primordial,puy2007primordial,Puy1993,Schleicher:2008,Lepp_2002,lepp1984molecules}.
According to papers above, the absence of dust and heavy elements in
the early Universe leads to extremely low molecular abundances. The
fraction of molecules in the baryonic gas in relation the cosmological
background reaches $\sim 10^{-5}-10^{-6}$ for hydrogen molecules,
$\sim 10^{-9}-10^{-10}$ for HD molecules, $\sim 10^{-11}-10^{-14}$ for
HeH$^+$ ion-molecules, and the fraction of LiH molecules reaches
$\sim 10^{-20}$ at the end of the Dark Ages. In the Dark Matter halos
chemical processes are expected to be more intense due to the denser
and hotter environment.  The number densities of H$_2$ and HD
molecules at the moment of halo virialization are estimated as
respectively $\sim 10^3$, it's $\sim 400$ times larger than that of
uniformly expanding background, while the number density of HeH$^+$
ion molecules on the contrary appears to be of orders of magnitude
lower \citep{Novosyadlyj2018, Novosyadlyj2022}. Despite quite low
abundances of primordial molecules in early Universe they are a
crucial ingredient for the formation of the first stars. Therefore,
their detection will provide an important insight into the processes
in early Universe that led to the emergence of the first stars,
galaxies, and supermassive black holes.

Dark Ages are studied by detecting spectral distortions of the CMB
caused by primordial chemistry and the 21 cm transition in neutral
hydrogen. The idea that resonant scattering of CMB photons in the
rovibrational lines of some primordial molecules of Dark Ages halos
could lead to an observable signature in the CMB thermal spectrum was
first proposed by \citet{Dubrovich1977} in the late 1970s. Since then,
a number of papers have analysed the possibility of detecting the
resonant lines (e.g. \citet{Maoli1994,
  Maoli1996,BASU2007431,DUBROVICH200828}) and there have been three
attempts to observe the effect, without success so far
\citep{Persson2010,Gosachinskij2002,de1993search}. Paper
\citet{Schleicher:2008} considers the distortion of
CMB spectrum due to absorption, photoionization, and
photo-dissociation processes during free-free and bound-free
transitions involving atomic and molecular ions on the cosmological
background of Dark Ages. There, most promising results were
obtained for the negative hydrogen ion H$^-$ and the HeH$^+$
ion-molecules. In this paper we have calculated the contribution to
the distortion of the CMB spectrum by bound-bound transitions between
rovibrational levels of the first molecules H$_2$, HD, and
HeH$^+$. LiH molecules are not considered due to their low abundance.

\section{Primordial chemistry} \label{sec:floats}

\begin{figure*}
\centerline{\includegraphics[width=0.7\textwidth]{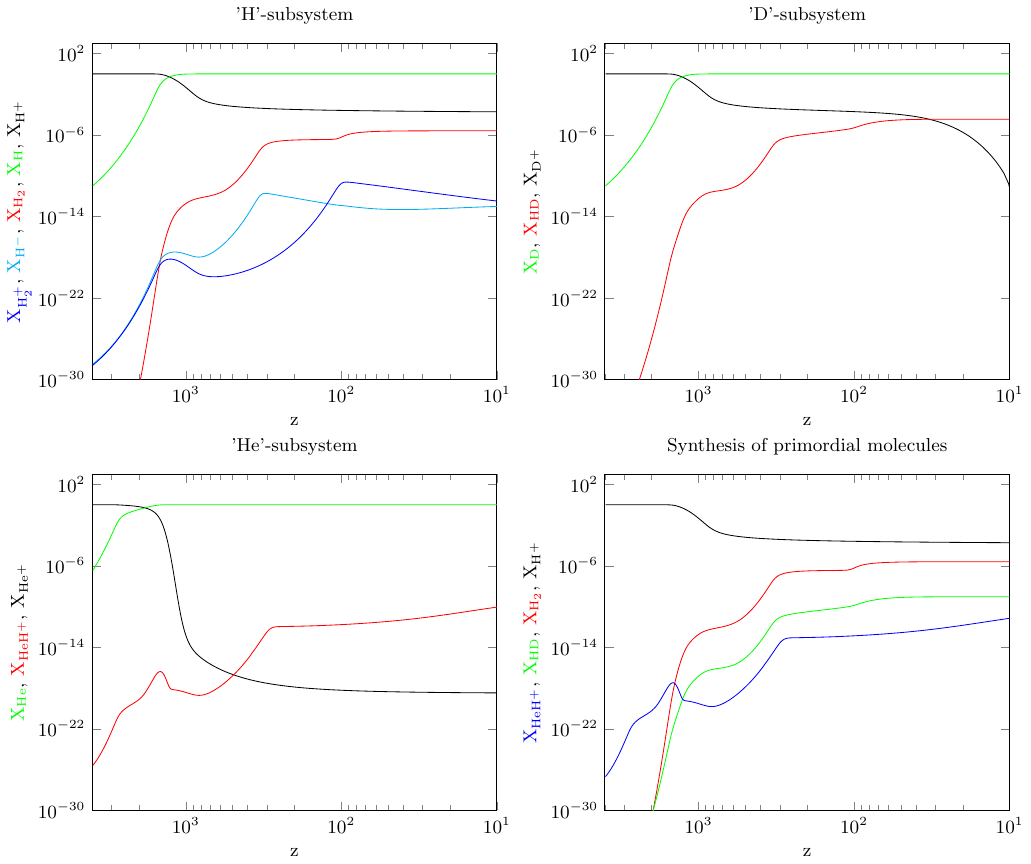} }
\caption{Abundances of atoms, molecules, and ions across redshifts.
Panels show relative abundances of key species for H$_2$ (top left), HD (top right), and HeH$^+$ (bottom left) in units of the density of hydrogen, deuterium, and helium nuclei, respectively. The bottom right panel displays the first molecules' abundances in hydrogen nucleus number density across redshifts.}
 \label{pr_chem}
\end{figure*}

Primordial chemical reactions start at the epoch of recombination when
neutral atoms emerged and continue in the post-recombination Universe
due to the presence of residuals of free electrons and ions. The
presence of such ingredients in almost neutral post-recombination gas
triggers a chain of chemical reactions, thus leading to the synthesis
of the first diatomic compounds, among which the most common are
molecules H$_2$ and HD, as well as the helium hydride ion
HeH$^+$. They are expected to have played a crucial role in forming
the first stars by allowing protostellar gas clouds to cool and
collapse. However, as mentioned above the primordial gas was
chemically depleted from atoms in various ionized and excited
states. In our calculations, we took into account twenty most common
reactants, namely photons of cosmic relic radiation ($\gamma$), e$^-$,
H, H$^+$, H$^-$, H$_2$, H$_2^+$, H$_3^+$, He, He$^{+}$, He$^{++}$,
HeH$^+$, D, D$^-$, D$^+$, D$_2$, D$_2^+$, HD, HD$^{+}$, and
H$_2$D$^{+}$. For these reagents, we consider 166 reactions listed in
Table~\ref{tab:long} subdivided into three groups: reactions not
involving He and D nuclei are labelled with the prefix ``H\_'',
reactions involving He nuclei and not involving D nuclei are labelled
with the prefix ``He\_'', reactions involving D nuclei are labelled
with the prefix ``D\_''. The Table shows the rates of the reactions as
a function of gas temperature, electron temperature, or radiation
temperature. The description in the Table lists the publications where
corresponding approximate expressions were obtained.

In the Table~\ref{tab:long} we present new approximate expressions for
the reaction rates of \hyperlink{H_21}{$\rm H_{21}$} and
\hyperlink{He_19}{$\rm He_{19}$}. The first one is calculated based on
the cross-section from the online database\footnote{
\url{https://home.strw.leidenuniv.nl/~ewine/photo/}}.  The second is
our approximation to the combined data from \citet{Schauer1989} and
\citet{Johnsen1980}. Some reactions involving molecules are sensitive
to their quantum state. For them, the Table~\ref{tab:long} contains
reaction rates for two cases: $v=0$ corresponds to the case when the
molecule is in the ground vibrational state, LTE corresponds to
condition of local thermodynamic equilibrium for the population of
rovibrational levels. Under conditions of early Universe the
population of rotational and vibrational levels of molecules is mainly
determined by the temperature of CMB radiation. At the same time, the
frequency of collisions depends on the temperature of the
gas. At redshifts $z>200$, the gas and background radiation
temperatures are equal, so it is correct to use LTE approximation. At
lower redshifts, the gas temperature is slightly below the temperature
of the background radiation, so the LTE approximation is less accurate
though we continue to use it.

Since the density of the $i$-th reactant, $n_{\rm i}$, changes not
only due to chemical reactions but also due to the expansion of the
Universe, it is convenient to represent the amount of the reactant in
the form of a fraction, that is, relative to the total number of
hydrogen nuclei as $x_{\rm i} \equiv n_{\rm i}/n_{\rm H}$. The
evolution of the fraction of the $i$-th reactant, $x_{\rm i}$, is
determined only by chemical reactions and is described by a kinetic
equation that has the following general form
\citep{Puy1993,Galli1998,Vonlanthen2009,Novosyadlyj2018}:
\begin{equation}
 \left(\frac{dx_i}{dt}\right)_{\rm{chem}}=\sum_{mn}k^{(i)}_{mn}n_{\rm H}x_{m}x_{n}+\sum_{m}k^{(i)}_{m\gamma}x_{m}
-\sum_{j}k_{ij}n_{\rm H}x_{i}x_{j}-k_{i\gamma}x_{i},
\label{chem_eqs}
\end{equation}
where {$k_{mn}^{(i)}$} is the reaction rate of reactants {$m$} and {$n$} with the formation of reactant {$i$}.
The number of differential equations can be reduced by applying
four equalities $1 = x_{\rm H I} + x_{\rm H II} + x_{\rm H^-} + x_{\rm HD} + x_{\rm HD^+} + x_{\rm HeH^+} + 2x_{\rm H_2^+} + 2x_{\rm H_2} + 2x_{\rm H_2D^+} + 3x_{\rm H_3^+}$,
$x_{\rm D} = x_{\rm D I} + x_{\rm D II}+ x_{\rm HD} + x_{\rm HD^+} + x_{\rm H_2D^+}$,
$x_{\rm He} = x_{\rm He I} + x_{\rm He II} + x_{\rm He II}+ x_{\rm HeH^+}$, and $x_e   = x_{\rm H II} + x_{\rm D II} + x_{\rm He II} + 2x_{\rm He III} + x_{\rm H_2^+}+ x_{\rm D_2^+} + x_{\rm HeH^+}  + x_{\rm H_3^+} + x_{\rm HD^+} + x_{\rm H_2D^+} - x_{\rm H^-}$.

In practice, we used the last equality to calculate the fraction of
free electrons, while the others allowed us to control the accuracy of
the integration of the differential equations. We did not calculate
the change in the number of photons since, as noted above, the number
of relic photons exceeds the number of photons arising from
recombination by nine orders of magnitude. All calculations were
performed for the period from the cosmological recombination up to
hydrogen reionization at $z_{rei}\approx 6$ and for corresponding values
of main parameters of the cosmological model as determined by final
data release of the Planck Space Observatory \citep{Planck2020},
listed in the previous section.

Main results of our calculations are shown in Fig.~\ref{pr_chem},
where the relative abundances of representative species involved in
the formation of H2 (top left panel), HD (top right), and HeH+ (bottom
left panel) are given in units of hydrogen nucleus number density,
deuterium nucleus number density, and helium nucleus number density,
respectively. The relative abundances of first molecules in units
of hydrogen nucleus number density are presented in the bottom right
panel. Our results qualitatively reproduce the results of other
authors, since the core of the most important reactions is common.

We have analysed chemical reactions for their influence on the
chemistry of the early Universe from the moment of recombination to
the moment of reionization. In the Table~\ref{tab:long}, we have
highlighted in bold those reactions that, in the indicated redshift
range, give a maximum contribution to the synthesis or destruction of
any of the reactants of more than 0.1 percent. Such an analysis shows
which of the reactions played a noticeable role in the early Universe
(within the framework of the standard cosmological model). It is
assumed that in halos and cosmological models with additional sources
of ionization and heating (primordial magnetic fields, decaying or
annihilating dark matter particles, primordial black holes) this set
will be somewhat different. The interactive chemical reaction map we
propose (i.e. Table~\ref{tab:1}) is a new way of visualizing a
network of chemical reactions, allowing us to visually track which
reactions are already included in the model of primordial chemistry,
and which ones can or should still be included.

\subsection{The Chemistry of H$_2$}

\begin{figure}
\centerline{\includegraphics[width=0.5\textwidth]{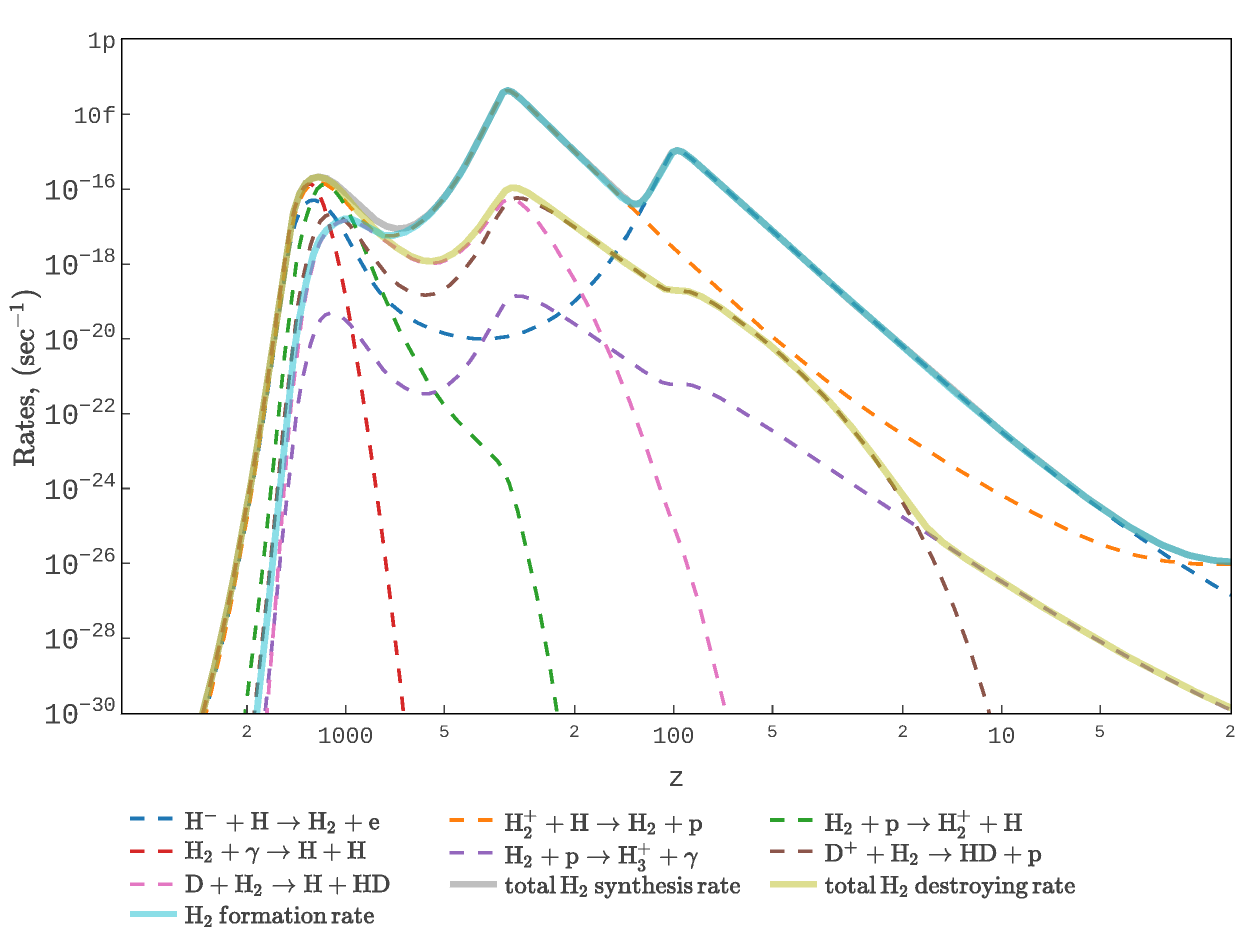} }
\caption{Kinetics of formation and destruction of H$_2$ molecules.
  The dependencies of the rates of the main reactions involved in the formation and destruction of H$_2$ molecules are provided.}
\label{gr:H2_formation}
\end{figure}

Direct radiative association of two neutral hydrogen atoms in the
ground states to form a hydrogen molecule is very unlikely (rate
$\ll 10^{-25}$~cm$^3$s$^{-1}$ \citep{Gould1963}) since the emitted
photon in this process must be a result of forbidden transition
between the two Hitler-London states of the molecule. Direct
radiative association of two neutral hydrogen atoms is possible if one
of the two colliding atoms is in an excited state (say, in the 2p
state), then the emission of photon will occur through an allowed
transition between the two states of the molecule. However, the
fraction of H atoms in the 2p and other excited states in the
post-recombination Universe is so small that this process can be
neglected. Therefore, in the conditions of the early Universe, two
indirect channels of formation of hydrogen molecules dominate -- one
through the mediation of ion $\rm H^{-}$:
\begin{equation}
 \rm{ H^- + H \to H_2 + e^-},\label{H2_01}
\end{equation}
and other through the mediation of ion-molecule  $\rm H_2^{+}$:
\begin{equation}
 \rm{H_2^+ + H \to H_2 + H^+}.\label{H2_02}
\end{equation}

As shown in Fig.\ref{gr:H2_formation}, the former process dominates at
$z \leqslant 127$ and reaches the maximum rates at the two peaks
located at $z\approx 1270$ and $z\approx 98$. The second process
dominates at $129\leqslant z \leqslant 1400$, reaching peak efficiency
at $z\approx 1220$ and $z\approx 320$. The gray and yellow lines in
Fig.\ref{gr:H2_formation} show the overall rates of synthesis and
destruction of H$_2$ molecules, respectively, which takes into account
all the involved reactions listed in Table~\ref{tab:long}. The total
formation rate is the difference between the synthesis and destruction
rates as shown in Fig.~\ref{gr:H2_formation} by the solid cyan
line. As can be seen from the figure, the formation of H$_2$ molecules
at $z \geqslant 1050$ is an equilibrium process, while at lower
redshifts, the rate of synthesis of H$_2$ molecules begins to prevail
over their destruction.

The main channel for creating $\rm H^{-}$
ions is the process
\begin{equation}
 \rm{ H + e^{-} \to H^{-} + \gamma}.\label{H2_03}
\end{equation}
The main channel for destruction $\rm H^{-}$ ions at $z \geqslant 100$ is process
\begin{equation}
 \rm{ H^{-} + \gamma \to H + e^{-}},\label{H2_04}
\end{equation}
while at $z \leqslant 100$, $\rm H^{-}$ ions are mainly spent on the
formation of the H$_2$ molecule during the process
(\ref{H2_01}). Since the formation of $\rm H^{-}$ ions is an
equilibrium process in the early Universe, their number density can be
easily estimated by the equality
$k_{(4)}n_{\rm H}n_{\rm e^{-}} = n_{\rm H^{-}}(k_{(5)}+k_{(2)}n_{\rm
  H})$, where $k_{(4)}$, $k_{(2)}$, and $k_{(3)}$ mark rates of
processes (\ref{H2_03}), (\ref{H2_01}), and (\ref{H2_04}),
respectively.

There are two main processes for formation of H$_2^+$, the first of
which
\begin{equation}
 \rm{ H + H^{+} \to H_2^{+} + \gamma},\label{H2p_07}
\end{equation}
dominates at  $z \leqslant 140$, while the second
\begin{equation}
 \rm{ HeH^{+} + H \to  H_2^{+} + He},\label{H2p_He04}
\end{equation}
dominates at $z \geqslant 140$. There are also two main processes for
destruction of H$_2^+$, the first of which
\begin{equation}
 \rm{H_2^{+} + \gamma \to H + H^{+}},\label{H2p_08}
\end{equation}
dominates at $z \leqslant 325$, while the second
\begin{equation}
 \rm{ H_2^{+} + H \to H_2 + H^{+}},\label{H2p_09}
\end{equation}
dominates at $z \geqslant 325$. The formation of $\rm H_2^{+}$ is an
equilibrium process in the early Universe, and their number density
can be easily estimated by the equality
$k_{(6)}n_{\rm H}n_{\rm H^{+}} + k_{(7)}n_{\rm HeH^{+}}n_{\rm H} =
n_{\rm H_2^{+}}(k_{(8)}+k_{(9)}n_{\rm H})$, where $k_{(6)}$,
$k_{(7)}$, $k_{(8)}$, and $k_{(9)}$ mark rates of processes
(\ref{H2p_07}), (\ref{H2p_He04}), (\ref{H2p_08}), and (\ref{H2p_09}),
respectively.

\subsection{The Chemistry of HD}

\begin{figure}
\centerline{\includegraphics[width=0.5\textwidth]{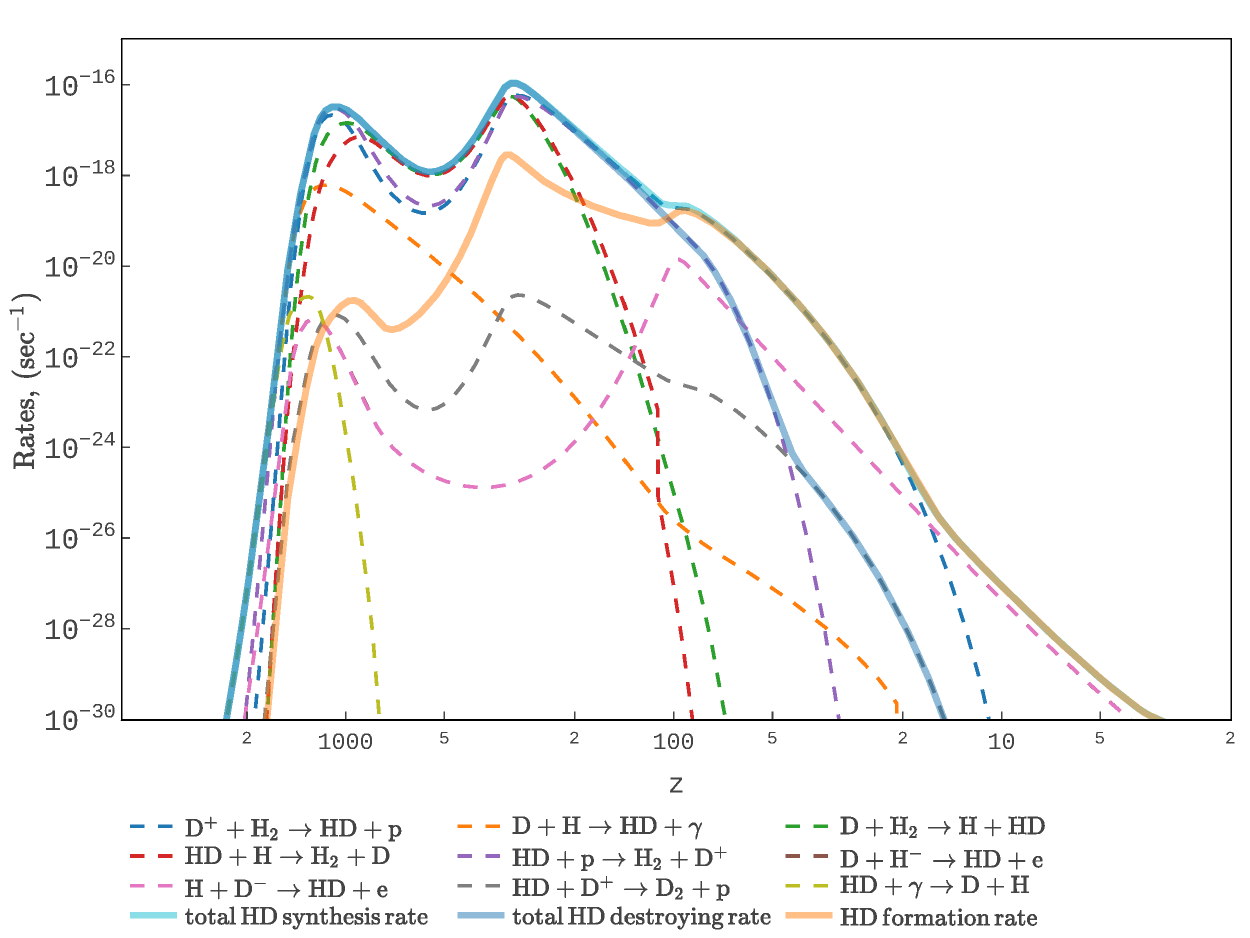} }
\caption{ Kinetics of formation and destruction of HD molecules. The
  dependencies of the rates of the main reactions involved in the
  formation and destruction of HD molecules are given.}
\label{gr:HD_formation}
\end{figure}

Unlike H$_2$, molecule of HD can be formed by direct radiative
association of two neutral atoms in the process
\begin{equation}
\rm H + D \to HD + \gamma,
\end{equation}
which reaches its maximum efficiency in the epoch of recombination
($z \approx 1100$). However, two processes of formation of HD
molecules from H$_2$ molecules are more effective in the conditions of
the early Universe, namely charge transfer
\begin{equation}
\rm H_2 + D^+ \to HD + H^+,\label{HD_01}
\end{equation}
and deuteron exchange
\begin{equation}
\rm H_2 + D \to HD + H.\label{HD_02}
\end{equation}
These processes have two peaks in efficiency at $z\approx 1180$ and
$z\approx 307$ and dominate at $z \geqslant 20$. At a later time, as
well as in the recombination epoch, processes involving H$^-$ and
D$^-$ ions should also be considered, namely
\begin{equation}
\rm H + D^- \to HD + e^-,\label{HD_03}
\end{equation}
and
\begin{equation}
\rm H^- + D \to HD + e^-.\label{HD_04}
\end{equation}

The cyan and blue lines in Fig.\ref{gr:HD_formation} show the overall
rates of synthesis and destruction of HD molecules when
all reactions from Table~\ref{tab:long} are involved. The total
formation rate is the difference between the synthesis and destruction
rates and is shown in Fig.~\ref{gr:HD_formation} by the solid orange
line. As it follows from Fig.~\ref{gr:HD_formation}, the formation of
HD molecule is an equilibrium process in early Universe at
$z \geqslant 120$. At later stages of evolution
($z \leqslant 120$), the synthesis of HD molecules dominates over
their destruction.

\subsection{The Chemistry of HeH$^+$}

\begin{figure}
\centerline{\includegraphics[width=0.5\textwidth]{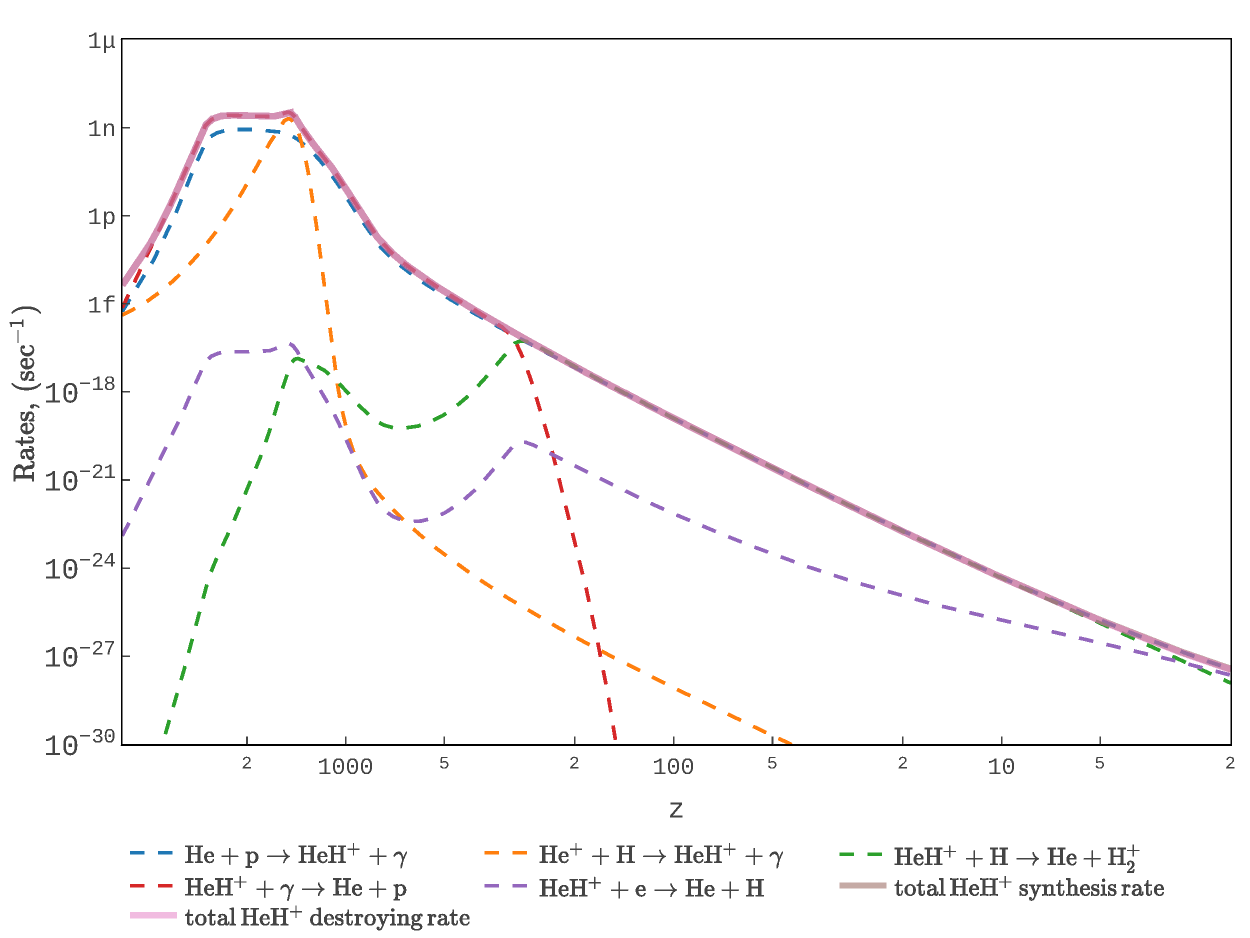} }
\caption{ Kinetics of formation and destruction of HeH$^+$
  ion-molecules. The dependencies of the rates of the main reactions
  involved in the formation and destruction of HeH$^+$ ion-molecules
  are given.}
\label{gr:HeHp_formation}
\end{figure}

As can be seen from Fig.~\ref{pr_chem} (see lower right panel), the
helium hydride ion (HeH$^+$) was the first molecular ion that began to
form in the early Universe during the recombination of He nuclei
during the recombination epoch due to three processes
\begin{equation}
\rm He + H^+ \to HeH^+ + \gamma,\label{HeH_1}
\end{equation}
and
\begin{equation}
\rm He^+ + H \to HeH^+ + \gamma,\label{HeH_2}
\end{equation}
and
\begin{equation}
\rm He^+ + H + \gamma \to HeH^+ + 2\gamma.\label{HeH_3}
\end{equation}

The main channel for destruction $\rm HeH^+$ ion-molecules at $z \geqslant 300$ is process
\begin{equation}
 \rm{ HeH^+ + \gamma \to He + H^{+}},\label{HeH_4}
\end{equation}
while at $z \leqslant 300$, $\rm HeH^+$ molecular ions were destroyed in collisions with neutral atoms of hydrogen due to process
\begin{equation}
 \rm{ HeH^+ + H \to He + H_2^{+}}.\label{HeH_5}
\end{equation}

The brown and pink lines in Fig.\ref{gr:HeHp_formation} show the
overall rates of synthesis and destruction of HeH$^+$ molecules,
taking into account all reactions listed in Table~\ref{tab:long}. The
fact that these lines merge in Fig.~\ref{gr:HeHp_formation} means that
the formation of HeH$^+$ molecules is an equilibrium process at all
redshifts, and their number density can be estimated by the equality
$k_{(15)}n_{\rm He}n_{\rm H^{+}} + (k_{(16)}+k_{(17)})n_{\rm
  He^{+}}n_{\rm H} = n_{\rm HeH^{+}}(k_{(18)}+k_{(19)}n_{\rm H})$,
where $k_{(15)}$, $k_{(16)}$, $k_{(17)}$, $k_{(18)}$, and $k_{(19)}$
mark rates of processes (\ref{HeH_1}), (\ref{HeH_2}), (\ref{HeH_3}),
(\ref{HeH_4}), and (\ref{HeH_5}), respectively.

\section{Rovibrational levels occupation} \label{sec:floats}

The kinetic equations for the population of rotational and vibrational
levels are as follows
\begin{eqnarray}
 \frac{dX_{v,j}^m}{dt} &=& \sum\limits_{s} n_{s}\sum\limits_{v',j'}\left[k^{s}_{v',j'\to v,j}X_{v',j'}^m-X_{v,j}^mk^{s}_{v,j\to v',j'}\right] \nonumber\\
 && + \left[\alpha_{v,j}^m\frac{d_{+}}{dt}-X_{v,j}^m\frac{d_{-}}{dt}\right]\ln(n_{m}a^3),
 \label{levels_occupation}
\end{eqnarray}
where $X^m_{v,j}=n_{m,\{v,j\}}/\sum_{v',j'}n_{m,\{v',j'\}}$ is the
fraction of molecules $m\in\{$H$_2$, HD, HeH$^+\}$ that is in the
quantum state $\{v,j\}$, and the index $s \in\{\gamma$, H, p, e$\}$
indicates the collision partner of molecule $m$ that caused the
transitions. Here, we assume that the transitions between the
rovibrational levels of Dark Age and Cosmic Down molecules are mainly
due to the emission/absorption of photons, collisions with neutral
hydrogen atoms, protons (responsible for
ortho$\leftrightarrows$para transitions in H$_2$) and electrons (
considered only for HeH$^+$). The last term of
Eq.~\ref{levels_occupation} describes the trend that population of
rovibrational levels of newly synthesized molecules, $\alpha_{v,j}^m$,
differs from the population of levels of molecules that have already
been exposed to the environment, $X_{v,j}^m$. The former is obviously
determined by the gas temperature,
\begin{equation*}
 \alpha_{v,j}^m \approx {g_j\exp\left\{-E_{v,j}/k_bT\right\}}/{\sum_{v',j'}g_{j'}\exp\left\{-E_{v',j'}/k_BT\right\}},
\end{equation*}
while the other is preferably determined by the CMB temperature.  The
derivative\footnote{The complete differential $d = d_+ - d_-$, where
  $d_+$ indicates an increase in the number of molecules due to their
  synthesis, and $d_-$ means the destruction of molecules due to their
  destruction.} $d_{+}\ln(n_ma^3)/dt$ specifies the rate of molecule
synthesis, while derivative $d_{-}\ln(n_ma^3)/dt$ determines the rate
of their destruction.

\begin{table}[tp]
\begin{center}
  \caption{Values of spontaneous rovibrational transitions allowed
    by quantum selection rules, along with frequencies and energy for the lowest rovibrational energy levels of para- and ortho-hydrogen
    molecules H$_2$, hydrogen deuteride molecules HD, and helium
    hydride ion HeH$^+$.}
\tiny
\begin{tabular} {ccccccc}
\hline
\hline
   \noalign{\smallskip}
Species&Transitions\footnote{Here and in the text $u$ means upper level, $l$ means lower level}&A$_{ul}$&  Frequency $\nu_{ul}$& E$_u$\\
 \noalign{\smallskip}
       &$j_u$\,--\,$j_l$ &[s$^{-1}$] & [GHz]      & [K] \\
       &$(v_u,j_u)$\,--\,$(v_l,j_l)$ & & & \\
  \noalign{\smallskip}
\hline
   \noalign{\smallskip}
ortho--H$_2$&1\,--\,$\,\,\,$ & -- &  -- & 170 \\
     &3\,--\,1&4.76\x10$^{-10}$	&17\,594&1\,015 \\
     &5\,--\,3&9.83\x10$^{-9}$	&31\,011&2\,503 \\
     &7\,--\,5&5.88\x10$^{-8}$  &43\,388&4\,586 \\
   \noalign{\smallskip}
para--H$_2$&2\,--\,0&2.94\x10$^{-11}$	&10\,621& 510   \\
     &4\,--\,2&2.75\x10$^{-9}$ 	&24\,410&1\,681\\
     &6\,--\,4&2.64\x10$^{-8}$	&37\,357&3\,475 \\
     &8\,--\,6&1.14\x10$^{-7}$	&49\,077&5\,830 \\
   \noalign{\smallskip}
HD   &(0,1)\,--\,(0,0)&5.41\x10$^{-8}$ 	&2\,675 &128   \\
     &(0,2)\,--\,(0,1)&5.13\x10$^{-7}$	&5\,332 &384    \\
     &(0,3)\,--\,(0,2)&1.80\x10$^{-6}$	&7\,952 &766    \\
     &(0,4)\,--\,(0,3)&4.31\x10$^{-6}$	&10\,518&1\,271 \\
     &(0,5)\,--\,(0,4)&8.35\x10$^{-6}$	&13\,015&1\,895 \\
     &(0,6)\,--\,(0,5)&1.40\x10$^{-5}$	&15\,429&2\,636 \\
     &(0,7)\,--\,(0,6)&2.14\x10$^{-5}$	&17\,746&3\,487 \\
     &(0,8)\,--\,(0,7)&3.03\x10$^{-5}$	&19\,955&4\,445 \\
     &(1,0)\,--\,(0,1)&3.09\x10$^{-5}$	& 106\,213& 5\,226 \\
     &(1,1)\,--\,(0,2)&1.55\x10$^{-5}$	&103\,441&5\,349 \\
     &(1,1)\,--\,(0,0)&1.67\x10$^{-5}$	&111\,447&5\,349 \\
     &(0,9)\,--\,(0,8)&4.07\x10$^{-5}$	& 22\,055&5\,503 \\
     &(1,2)\,--\,(0,1)&2.46\x10$^{-5}$	& 113\,874 & 5\,594\\
     &(1,2)\,--\,(0,3)&1.01\x10$^{-5}$	& 100\,590 & 5\,594\\
     &(1,3)\,--\,(0,2)&3.16\x10$^{-5}$	& 116\,150 & 5\,959\\
     &(1,3)\,--\,(0,4)&6.73\x10$^{-6}$	& 97\,679 & 5\,959\\
     &(1,4)\,--\,(0,3)&3.86\x10$^{-5}$	& 118\,259 & 6\,442\\
     &(1,4)\,--\,(0,5)&4.34\x10$^{-6}$	& 94\,724 & 6\,442\\
     &(0,10)\,--\,(0,9)&5.23\x10$^{-5}$	& 24\,038&6\,657 \\

   \noalign{\smallskip}
HeH$^+$
     &1\,--\,0&0.109 	&2\,010 &96   \\
     &2\,--\,1&1.04 &4\,009 &289    \\
     &3\,--\,2&3.75 &5\,984 &576    \\
     &4\,--\,3&9.14 &7\,925 &956 \\
     &5\,--\,4&18.1 &9\,821 &1\,428 \\
 \noalign{\smallskip}
  \hline
\end{tabular}
\label{Tab2}
\end{center}
\end{table}

The main process to define the population of the rovibrational levels
of the first molecules in the early Universe was collisions with
quanta of cosmic microwave background radiation. Since this radiation
is a blackbody, the population of rovibrational levels can be
considered to be in equilibrium with great accuracy, i.e., described
by a Boltzmann distribution with temperature $T_\text{R}$. The
contributions of spontaneous and radiative rovibrational transitions
to the right-hand side of the equation (\ref{levels_occupation}) have
the following form
\begin{equation*}
 n_{\gamma}\cdot k^{\gamma}_{ul} = A_{ul} + B_{ul}\cdot U(\nu_{ul}),\quad
 n_{\gamma}\cdot k^{\gamma}_{lu} = B_{lu}\cdot U(\nu_{ul}),\nonumber
\end{equation*}
where index '$ul$' means up to low levels transition whereas index
'$lu$' means low to up levels transition, $A_{ul}$ is the rate of
spontaneous emission, $B_{ul} = (c^3/4h\nu^3_{ul})A_{ul}$ is the rate
of radiative emission, $B_{lu} = B_{ul}$ is the rate of radiative
absorption, and
\begin{equation*}
 U(\nu) = \frac{4h\nu^3}{c^3} \frac{1}{\exp\left(h\nu/kT_R\right)-1}
\end{equation*}
is the blackbody energy density. The values of transition frequencies
and energies for allowed spontaneous rovibrational transitions are
listed in Table~\ref{Tab2}, for low rovibrational energy levels. Data
for para- and ortho- hydrogen molecules H$_2$ are taken from
\cite{Roueff:2019}, data for hydrogen deuteride molecules HD and
helium hydride ion HeH$^+$ are taken from \cite{Amaral:2019}.

Spontaneous emission and radiative rovibrational transitions for
diatomic molecules with non-zero electric dipole moments (case for HD
and HeH$^+$) obey quantum selection rules $\Delta j = \pm 1$ and
$\Delta v = \pm 1$. Hydrogen molecules are homonuclear diatomic
molecules with a zero electric dipole moment. For this reason, it is
two types of them depending on the relative orientation of the nuclear
spins: ortho- and para-. Ortho-hydrogen molecules are those in which
the spins of both nuclei (protons) point in the same direction, so
their total spin is $I_n+I_n=1$, where $I_n=1/2$ is the spin of one
nucleus (proton). Molecules of hydrogen in which spins of both nuclei
are directing in opposite directions are called para-hydrogen, and
their total spin is $I_n+I_n=0$. Spontaneous emission and radiative
rovibrational transitions do not lead to the mixing of these spin
isomers of hydrogen molecules because they obey the quantum selection
rules $\Delta j = \pm 2$ and $\Delta v = \pm 1$. Spin isomers of
homonuclear diatomic molecules have distinct sequences of statistical
weights of rotational levels, $g_j=g_n(2j+1)$, which differ in nuclear
statistical weights -- $g_{n} = I_n(2I_n+1)$ or
$g_{n} = (I_n+1)(2I_n+1)$ depending on the parity of $j$ and
symmetries of the binding wave functions. Ortho-hydrogen corresponds
to odd values of $j$, and its nuclear statistical weight is $g_n=3$,
while for para-hydrogen, with even values of $j$, nuclear statistical
weight is $g_n=1$. The statistical weight of any vibrational level $v$
of a diatomic harmonic oscillator is $g_v = 1$.

The next factor affecting the population of the rovibrational levels
of molecules is their collisions with the components of the baryon
gas. As seen from Fig.~\ref{pr_chem}, neutral hydrogen atoms used to
be most abundant. In this paper, we used the collisional de-excitation
rates, $k_{ul}$, of rovibrational levels of H$_2$ molecules by H given
in \cite{Lique:2015}, where $u$ and $l$ mean upper and lower
rovibrational energy levels. The reverse transition rate (excitation)
coefficients can be obtained as follows
\begin{equation}
 k_{lu}(T) = \frac{g_u}{g_l}\exp\left\{-\frac{E_u-E_l}{k_{\rm B}T}\right\}k_{ul}(T),
\end{equation}
where $E_u$ and $E_l$ are the energies of the levels, $g_u$ and $g_l$
are statistical weights of the levels, and $T$ is the baryon gas
temperature. We used the rates from \cite{Desrousseaux:2022} for
collisional de-excitation of HD molecules by H. For collisional
de-excitation of HeH$^+$ molecules by H we used rates from
\cite{Kulinich:2020}, which are somewhat higher than those obtained in
\cite{Desrousseaux:2020}. Although collisions of molecules H$_2$ with
protons in the early Universe are much rarer, they nevertheless make a
significant contribution to ortho-para transitions:
$\rm p + H_2(j) \leftrightarrow p + H_2(j')$, where
$ \Delta j = j' - j = \pm 1, \pm 2, \pm 3, ...$. For collisional
de-excitation of H$_2$ molecules by $p$, we used the rates from paper
\cite{Gerlich:1990}. Free electrons are faster than free protons, so
their impact on the excitation and deactivation of the rotational
levels of ion molecules HeH$^+$ is more significant. To describe the
rotational transitions in the helium hydride ion, HeH$^+$, caused by
collisions with free electrons, we used the transition rates from
\cite{Khamesian:2018}, which unfortunately are only given for
$j\le 5$.

\section{Global signals of the first molecules} \label{sec:floats}

   \begin{figure}
   \centering
   \includegraphics[width=0.4\textwidth]{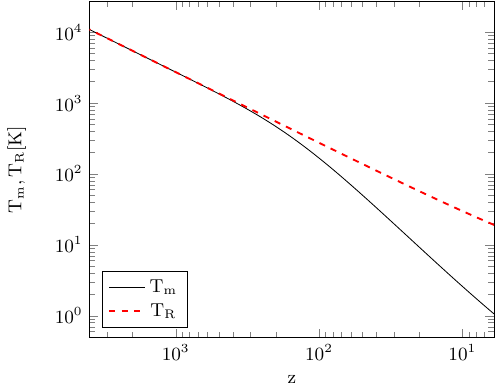}
   \caption{Thermal history of the early Universe.}
              \label{thermalhystory}
    \end{figure}

\begin{figure*}
   \centering
   \includegraphics[width=0.7\textwidth]{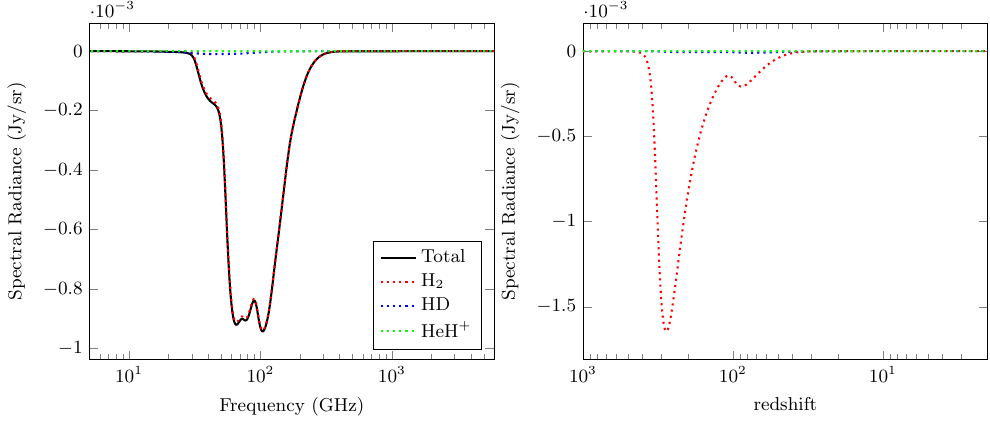}
   \caption{Left panel: superposition of absorption profiles of the
     first molecules H$_2$ and HD, and ion-molecule HeH$^+$. Right
     panel: redshifts at which absorption occurred.}
\label{totmolsyg}
\end{figure*}

During the Dark Ages, the temperature of baryonic matter is mainly
determined by the competition between adiabatic cooling and heating
due to the Compton scattering of CMBRs on free electrons.  Other
factors affecting the energy balance of the gas are collisional
activations and de-activations of molecular rovibrational levels and
primordial chemistry. However, as it shown in
\citet{Puy1993,Novosyadlyj:2023}, these processes make a small
contribution to the total energy balance of the gas in the
cosmological background.  Thus, energy balance for the baryonic
component on the cosmic background can be expressed by the equation of
the evolution of its temperature with time
\begin{equation}
 \frac{d T_m}{dt} = -2HT_m + \frac{8\sigma_TaT_{\rm R}^4(T_{\rm R}-T_m)x_e}{3m_ec},
\end{equation}
where $\sigma_T$ is the Thomson cross-section, $a$ is the radiation
constant, $m_e$ is the electron mass, and $x_e$ is the ionization
fraction, $n_b$ -- is baryonic gas number density.  Energy exchange
between free electrons and the radiation field played a leading role
in the early Universe, causing the equalization of baryon gas and
radiation temperatures at redshifts $z>300$ (see
Fig.~\ref{thermalhystory}). At lower redshifts, the transfer of energy
from radiation to the baryonic gas slows down significantly due to the
reduced frequency of collisions between electrons and quanta of CMB
radiation in the expanding Universe. From this moment on, adiabatic
cooling begins to dominate, therefore baryon gas cooled
faster than radiation.

Population of the rovibrational levels starts to deviate from the
equilibrium at redshifts $z<300$ due to collisions between molecules
and other components of the baryon gas, where the temperature of the
baryon gas is slightly lower than the temperature of the CMB
radiation. These deviations can be represented as a deviation of the
excitation temperature,
\begin{equation}
 T^{ul}_{ex} \equiv \frac{E_u-E_l}{k_B}\ln^{-1}\left(\frac{g_un_l}{g_ln_u}\right),
\end{equation}
from the radiation temperature $T_R$. The differential brightness of
the first molecules on the background of CMB can be expressed as
\begin{equation*}
 \delta I_{ul} = \frac{2h\nu^3_{ul}}{c^2}\left[\frac{1}{e^{\frac{h\nu_{ul}}{k_{B}T_{ex}}}-1} - \frac{1}{e^{\frac{h\nu_{ul}}{k_{B}T_{r}}}-1}\right]\tau_{ul},
\end{equation*}
where the optical depth in the transition lines $u\to l$ in the
approximation of narrow line looks as follows
\begin{equation*}
\tau_{ul} = \frac{1}{8\pi} \frac{\lambda_{ul}^3A_{ul}}{H(z)}n_{{\rm mol}}\left(X_{l}\frac{g_{u}}{g_{l}}-X_{u}\right),
\end{equation*}
where $\lambda_{ul}=c/\nu_{ul}$ is wavelength, $H(z)$ is expansion
rate that depend on redshift, $n_{\rm mol}$ is number density of
molecules.  Note, we can not observe individual lines but their
superposition
\begin{equation*}
\delta I_{\rm tot} = \sum\limits_{\rm transitions}\delta I_{ul}.
\end{equation*}

Differential brightness of the molecules in the rovibrational lines,
as well as their superposition, are shown in Fig.~\ref{mol_sygnals},
where the left panels show absorption profiles in the rotational lines
of molecules H$_2$ (upper panel) and HD (middle panel), and
ion-molecule HeH$^+$ (bottom panel). The dotted coloured lines show
contributions from separate transitions, whereas the black line shows
their superposition. The right panels on Fig.~\ref{mol_sygnals} show
the redshifts at which absorption occurred for corresponding
transitions for H$_2$ (upper panel) and HD (middle panel) molecules
and for ion-molecule HeH$^+$ (bottom panel). As you can see,
absorption begins at $z\simeq 300$ when the gas temperature $T$ starts
to deviate from the CMB temperature $T_R$.  The absorption profile of
H$_2$ molecule has a multi-peak feature and reaches a maximum
amplitude of $\sim 10^{-3}$ Jy/sr within the frequency range from
about 50 GHz to about 120 GHz. The maximum absorption in rovibrational
lines of H$_2$ molecules comes from redshifts $300>z>200$.  The
absorption profile of the HD molecule has a double peak feature and
reaches a maximum amplitude of $\sim 10^{-5}$ Jy/sr within the
frequency range of 40 GHz to 70 GHz. Maximal absorption in
rovibrational lines of HD molecules comes from redshifts $300>z>30$.
The absorption profile of HeH$^+$ ion-molecule has no features and
reaches a maximum amplitude of $\sim 10^{-7}$ Jy/sr within the
frequency range of 200 GHz to 800 GHz. The maximal contribution to
absorption in rovibrational lines of HeH$^+$ ion-molecule comes from
redshifts $100>z>4$.

The superposition of absorption profiles of the first molecules H$_2$
and HD, and ion-molecule HeH$^+$ is shown on the left panel of
Fig.~\ref{totmolsyg}. As you can see, despite zero electric dipole
moments of molecules H$_2$, their absorption signal dominates over the
HD absorption signal by two orders of magnitude and dominates over the
HeH$^+$ absorption signal by four orders of magnitude. This happens
due to a larger population of hydrogen molecules among all other
compounds in the early post-recombination Universe (see
Fig.~\ref{pr_chem}). From the right panel of Fig.~\ref{totmolsyg}, one
can see that molecular absorption occurs in the Dark Ages of the
Universe (lie in the redshift range $1000<z<30$).

\section{Discussion and conclusions}

The global signal from first molecules is a useful source of
information from yet undiscovered parts of the visible Universe. It
can provide information about the thermal and ionization history of
baryonic matter as well as the kinetics of primordial chemistry in the
epoch of the Dark Ages. In standard cosmology, this signal takes a
form of an absorption profile on the CMB spectrum.

Our estimates of the expected signal from H$_2$ and HD molecules are
of two orders of magnitude lower for the H$_2$ case and three orders of
magnitude lower for the HD case compared to the results of
\cite{Puy1993}. This discrepancy can be explained by the
difference in cosmological models. In the model we used, the baryon
abundance is only $\Omega_b = 0.0493$ in contrast to $\Omega_b = 1$
in \cite{Puy1993}. The difference in results was also influenced by
the difference in the models of primordial chemistry used. What is
similar to our results is the position of the H$_2$ and HD signals in
the CMB spectrum.

The effects of primordial chemistry on the CMB spectrum was previously
studied also in \cite{Schleicher:2008} where free-free and bound-free
absorption and emission processes were considered for ions H$^-$ and
He$^-$ as well as for ion-molecules HD$^+$ and HeH$^+$. There, most
promising results were obtained for the negative hydrogen ion H$^-$
for which the relative change in the CMB temperature lies in the range
$10^{-15} \lesssim \delta T_R/T_R \lesssim 10^{-6}$ for the frequency
range of $10\,\textrm{GHz} < \nu < 2000\,\textrm{GHz}$ (see Fig.~10 in
\cite{Schleicher:2008}).  The authors did not take into account
absorption by H$_2$ and HD molecules, considering it to be relatively
small.

To compare molecular absorption obtained here with the ionic
absorption from \cite{Schleicher:2008}, we use the formula that
relates small-amplitude intensity fluctuations to small-amplitude
temperature fluctuations:
\begin{equation*}
 \frac{\delta I}{I} \approx  f(x) \frac{\delta T_R}{T_R},
\end{equation*}
where $f(x) = x\cdot\exp(x)/(\exp(x)-1)$, and $x={h\nu}/{k_BT_R}$.  At
the frequency $\nu \approx 10^2$~GHz, where the absorption amplitude
of first molecules is at maximum
$\delta I \approx 0.9\cdot 10^{-3}$~Jy/sr (see Fig.~\ref{totmolsyg}),
the CMB specific intensity is equal to $I \approx 3\cdot
10^8$~Jy/sr. For $\nu = 10^2$~GHz and $T_R \approx 2.7$~K we obtain
$x \approx 1.7$ and $f(x) \approx 2.0$. Therefore, relative
fluctuations of the CMB specific intensity at frequency
$\nu = 10^2$~GHz is estimated at the level of
$\delta I/I \approx 3\cdot 10^{-12}$ in correspondence to relative
fluctuations of the CMB temperature at the level of
$\delta T_R/T_R \approx 1.5\cdot 10^{-12 }$. So the maximum of first
molecules signal is located in the minimum of the signal from negative
hydrogen ion (see Fig.~10 in \cite{Schleicher:2008}), exceeding it by
several orders of magnitude.  This means that molecular absorption
cannot be neglected in comparison to ionic absorption.

To assess the possibility of detecting a signal from the first
molecules, other sources of the CMB spectrum distortion in the
standard $\Lambda$CDM cosmology should be taken into account. These
are $\mu$- and $y$-distortions, a temperature shift due to a relative
error of the measured CMB temperature
$|\Delta T_R/T_R| \lesssim 5\times 10^{-4}$, and the cosmological
recombination radiation \citep{Desjacques:2015}. The frequency
dependencies of the first three are monotonic and well-determined so
that they can be removed from the overall signal. The latter depends
on the details of the recombination of H and He at $z\sim 10^3$ and
has a non-monotonic character with excess CMB emission at the
frequency of $\sim 10^{2}$~GHz at the level of $\sim 10^{-1}$~Jy/sr.
These are two orders of magnitude larger than the signal from the
first molecules.

Therefore, we conclude that detecting the signal from
the first molecules will require the accuracy of predicting the
cosmological recombination radiation up to the third order of
magnitude, as well as a radio telescope with a sensitivity of
$\sim 10^{-4}$~Jy/sr with spectral resolution of several tens of GHz.

\begin{acknowledgements}
  We thank an anonymous referee whose comments significantly improved
  this article.  This work is done in the framework of the project
  \emph{''Tomography of the Dark Ages and Cosmic Dawn in the lines of
    hydrogen and the first molecules as a test of cosmological
    models''} (state registration number 0124U004029) supported by
  National Research Found of Ukraine. The authors are grateful to
  Pavlo Kopach for technical assistance in the preparation of the
  paper.
\end{acknowledgements}

\bibliographystyle{aa}
\bibliography{refs}

\begin{thebibliography}{98}
\expandafter\ifx\csname natexlab\endcsname\relax\def\natexlab#1{#1}\fi

\bibitem[{{Adams} \& {Smith}(1985)}]{Adams1985}
{Adams}, N.~G. \& {Smith}, D. 1985, \apjl, 294, L63

\bibitem[{{Albertsson} {et~al.}(2014){Albertsson}, {Indriolo}, {Kreckel},
  {Semenov}, {Crabtree}, \& {Henning}}]{Albertsson2014}
{Albertsson}, T., {Indriolo}, N., {Kreckel}, H., {et~al.} 2014, \apj, 787, 44

\bibitem[{{Amaral} {et~al.}(2019){Amaral}, {Diniz}, {Jones}, {Stanke},
  {Alijah}, {Adamowicz}, \& {Mohallem}}]{Amaral:2019}
{Amaral}, P. H.~R., {Diniz}, L.~G., {Jones}, K.~A., {et~al.} 2019, \apj, 878,
  95

\bibitem[{Aver {et~al.}(2015)Aver, Olive, \& Skillman}]{Ave15}
Aver, E., Olive, K.~A., \& Skillman, E.~D. 2015, JCAP, 2015, 011

\bibitem[{Basu(2007)}]{BASU2007431}
Basu, K. 2007, New Astron. Rev., 51, 431, francesco Melchiorri: Scientist,
  Pioneer, Mentor

\bibitem[{{Black}(1978)}]{Black1978}
{Black}, J.~H. 1978, \apj, 222, 125

\bibitem[{Black(2006)}]{black2006chemistry}
Black, J.~H. 2006, Faraday Disc., 133, 27

\bibitem[{Bougleux \& Galli(1997)}]{bk978-0-7503-1425-1ch1bib6}
Bougleux, E. \& Galli, D. 1997, \mnras, 288, 638

\bibitem[{Bovino {et~al.}(2011b)Bovino, Tacconi, Gianturco, \&
  Galli}]{bk978-0-7503-1425-1ch1bib7}
Bovino, S., Tacconi, M., Gianturco, F.~A., \& Galli, D. 2011b, A\&A, 529, A140

\bibitem[{Bovino {et~al.}(2009)Bovino, Wernli, \& Gianturco}]{bovino2009fast}
Bovino, S., Wernli, M., \& Gianturco, F.~A. 2009, \apj, 699, 383

\bibitem[{{Coc} \& {Vangioni}(2017)}]{Coc17}
{Coc}, A. \& {Vangioni}, E. 2017, Int. J. Mod. Phys. E, 26, 1741002

\bibitem[{{Cooke} {et~al.}(2014){Cooke}, {Pettini}, {Jorgenson}, {Murphy}, \&
  {Steidel}}]{Coo14}
{Cooke}, R.~J., {Pettini}, M., {Jorgenson}, R.~A., {Murphy}, M.~T., \&
  {Steidel}, C.~C. 2014, \apj, 781, 31

\bibitem[{{Coppola} {et~al.}(2011){Coppola}, {Diomede}, {Longo}, \&
  {Capitelli}}]{Coppola2011b}
{Coppola}, C.~M., {Diomede}, P., {Longo}, S., \& {Capitelli}, M. 2011, \apj,
  727, 37

\bibitem[{{Coppola} {et~al.}(2017){Coppola}, {Kazandjian}, {Galli}, {Heays}, \&
  {van Dishoeck}}]{Coppola2017}
{Coppola}, C.~M., {Kazandjian}, M.~V., {Galli}, D., {Heays}, A.~N., \& {van
  Dishoeck}, E.~F. 2017, \mnras, 470, 4163

\bibitem[{Coppola {et~al.}(2011)Coppola, Longo, Capitelli, Palla, \&
  Galli}]{bk978-0-7503-1425-1ch1bib23}
Coppola, C.~M., Longo, S., Capitelli, M., Palla, F., \& Galli, D. 2011, \apjs,
  193, 7

\bibitem[{{Coppola} {et~al.}(2011){Coppola}, {Longo}, {Capitelli}, {Palla}, \&
  {Galli}}]{Coppola2011a}
{Coppola}, C.~M., {Longo}, S., {Capitelli}, M., {Palla}, F., \& {Galli}, D.
  2011, \apjs, 193, 7

\bibitem[{{Courtney} {et~al.}(2021){Courtney}, {Forrey}, {McArdle}, {Stancil},
  \& {Babb}}]{Courtney2021}
{Courtney}, E.~D.~S., {Forrey}, R.~C., {McArdle}, R.~T., {Stancil}, P.~C., \&
  {Babb}, J.~F. 2021, \apj, 919, 70

\bibitem[{{Croft} {et~al.}(1999){Croft}, {Dickinson}, \& {Gadea}}]{cdg99}
{Croft}, H., {Dickinson}, A.~S., \& {Gadea}, F.~X. 1999, \mnras, 304, 327

\bibitem[{Dalgarno {et~al.}(1996)Dalgarno, Kirby, \&
  Stancil}]{bk978-0-7503-1425-1ch1bib25}
Dalgarno, A., Kirby, K., \& Stancil, P.~C. 1996, ApJ, 458, 397

\bibitem[{{Dalgarno} \& {Lepp}(1987)}]{dl87}
{Dalgarno}, A. \& {Lepp}, S. 1987, in IAU Symposium, Vol. 120, Astrochem., ed.
  M.~S. {Vardya} \& S.~P. {Tarafdar}, 109--118

\bibitem[{{Dalgarno} \& {McDowell}(1956)}]{dm56}
{Dalgarno}, A. \& {McDowell}, M.~R.~C. 1956, Proc. Phys. Soc. A, 69, 615

\bibitem[{{Datz} {et~al.}(1995){Datz}, {Larsson}, {Stromholm}, {Sundstr{\"o}m},
  {Zengin}, {Danared}, {K{\"a}llberg}, \& {Ugglas}}]{Datz1995}
{Datz}, S., {Larsson}, M., {Stromholm}, C., {et~al.} 1995, \pra, 52, 2901

\bibitem[{de~Bernardis {et~al.}(1993)de~Bernardis, Dubrovich, Encrenaz, Maoli,
  Masi, Mastrantonio, Melchiorri, Melchiorri, Signore, \&
  Tanzilli}]{de1993search}
de~Bernardis, P., Dubrovich, V., Encrenaz, P., {et~al.} 1993, A\&A, 269, 1

\bibitem[{{De Fazio}(2014)}]{Fazio:2014}
{De Fazio}, D. 2014, PCCP, 16, 11662

\bibitem[{{Desjacques} {et~al.}(2015){Desjacques}, {Chluba}, {Silk}, {de
  Bernardis}, \& {Dor{\'e}}}]{Desjacques:2015}
{Desjacques}, V., {Chluba}, J., {Silk}, J., {de Bernardis}, F., \& {Dor{\'e}},
  O. 2015, \mnras, 451, 4460

\bibitem[{{Desrousseaux} {et~al.}(2022){Desrousseaux}, {Coppola}, \&
  {Lique}}]{Desrousseaux:2022}
{Desrousseaux}, B., {Coppola}, C.~M., \& {Lique}, F. 2022, \mnras, 513, 900

\bibitem[{{Desrousseaux} \& {Lique}(2020)}]{Desrousseaux:2020}
{Desrousseaux}, B. \& {Lique}, F. 2020, \jcp, 152, 074303

\bibitem[{{Dickinson}(2005)}]{dic05}
{Dickinson}, A.~S. 2005, J. Phys. B Atom. Mol. Phys., 38, 4329

\bibitem[{{Dove} {et~al.}(1987){Dove}, {Rusk}, {Cribb}, \& {Martin}}]{drcm87}
{Dove}, J.~E., {Rusk}, A.~C.~M., {Cribb}, P.~H., \& {Martin}, P.~G. 1987, \apj,
  318, 379

\bibitem[{Dubrovich {et~al.}(2008)Dubrovich, Bajkova, \&
  Khaikin}]{DUBROVICH200828}
Dubrovich, V., Bajkova, A., \& Khaikin, V. 2008, New Astron., 13, 28

\bibitem[{Dubrovich(1977)}]{Dubrovich1977}
Dubrovich, V.~K. 1977, Sov. Astron. Lett., 3, 128

\bibitem[{{Ferland} {et~al.}(1992){Ferland}, {Peterson}, {Horne}, {Welsh}, \&
  {Nahar}}]{FER92}
{Ferland}, G.~J., {Peterson}, B.~M., {Horne}, K., {Welsh}, W.~F., \& {Nahar},
  S.~N. 1992, \apj, 387, 95

\bibitem[{{Galli} \& {Palla}(1998)}]{Galli1998}
{Galli}, D. \& {Palla}, F. 1998, \aap, 335, 403

\bibitem[{Galli \& Palla(2002)}]{galli2002deuterium}
Galli, D. \& Palla, F. 2002, P\&SS, 50, 1197

\bibitem[{Gay {et~al.}(2011)Gay, Stancil, Lepp, \&
  Dalgarno}]{bk978-0-7503-1425-1ch1bib38}
Gay, C.~D., Stancil, P.~C., Lepp, S., \& Dalgarno, A. 2011, ApJ, 737, 44

\bibitem[{{Gerlich}(1990)}]{Gerlich:1990}
{Gerlich}, D. 1990, \jcp, 92, 2377

\bibitem[{{Glover} \& {Abel}(2008)}]{Glover2008}
{Glover}, S.~C.~O. \& {Abel}, T. 2008, \mnras, 388, 1627

\bibitem[{Gosachinskij {et~al.}(2002)Gosachinskij, Dubrovich, Zhelenkov, Il'in,
  \& Prozorov}]{Gosachinskij2002}
Gosachinskij, I.~V., Dubrovich, V.~K., Zhelenkov, S.~R., Il'in, G.~N., \&
  Prozorov, V.~A. 2002, Astron. Rep., 46, 543

\bibitem[{{Gould} \& {Salpeter}(1963)}]{Gould1963}
{Gould}, R.~J. \& {Salpeter}, E.~E. 1963, \apj, 138, 393

\bibitem[{{Huq} {et~al.}(1982){Huq}, {Doverspike}, {Champion}, \&
  {Esaulov}}]{h82}
{Huq}, M.~S., {Doverspike}, L.~D., {Champion}, R.~L., \& {Esaulov}, V.~A. 1982,
  J. Phys. B Atom. Mol. Phys., 15, 951

\bibitem[{{Janev} {et~al.}(1987){Janev}, {Langer}, {Post}, \& {Evans}}]{JAN87}
{Janev}, R.~K., {Langer}, W.~D., {Post}, Douglas~E., J., \& {Evans}, Kenneth,
  J., eds. 1987, {Elementary processes in hydrogen-helium plasmas: Cross
  sections and reaction rate coefficients}, Vol.~4

\bibitem[{Janev {et~al.}(2003)Janev, Reiter, \& Samm}]{Janev2003}
Janev, R.~K., Reiter, D., \& Samm, U. 2003, Collision Processes in
  Low-Temperature Hydrogen Plasmas, Tech. Rep. Jül-4105, Forschungszentrum
  Jülich GmbH, Jülich, Germany

\bibitem[{{Johnsen} {et~al.}(1980){Johnsen}, {Chen}, \& {Biondi}}]{Johnsen1980}
{Johnsen}, R., {Chen}, A., \& {Biondi}, M.~A. 1980, \jcp, 72, 3085

\bibitem[{{Karpas} {et~al.}(1979){Karpas}, {Anicich}, \& {Huntress}}]{KAR79}
{Karpas}, Z., {Anicich}, V., \& {Huntress}, W.~T. 1979, \jcp, 70, 2877

\bibitem[{{Khamesian} {et~al.}(2018){Khamesian}, {Ayouz}, {Singh}, \&
  {Kokoouline}}]{Khamesian:2018}
{Khamesian}, M., {Ayouz}, M., {Singh}, J., \& {Kokoouline}, V. 2018, Atoms, 6,
  49

\bibitem[{{Kimura} {et~al.}(1993){Kimura}, {Lane}, {Dalgarno}, \&
  {Dixson}}]{kldd93}
{Kimura}, M., {Lane}, N.~F., {Dalgarno}, A., \& {Dixson}, R.~G. 1993, \apj,
  405, 801

\bibitem[{{Kulinich} {et~al.}(2020){Kulinich}, {Novosyadlyj}, {Shulga}, \&
  {Han}}]{Kulinich:2020}
{Kulinich}, Y., {Novosyadlyj}, B., {Shulga}, V., \& {Han}, W. 2020, \prd, 101,
  083519

\bibitem[{{Larsson} {et~al.}(1996){Larsson}, {Lepp}, {Dalgarno}, {Stroemholm},
  {Sundstroem}, {Zengin}, {Danared}, {Kaellberg}, {Af Ugglas}, \&
  {Datz}}]{Larsson1996}
{Larsson}, M., {Lepp}, S., {Dalgarno}, A., {et~al.} 1996, \aap, 309, L1

\bibitem[{Latter \& Black(1991)}]{bk978-0-7503-1425-1ch1bib54}
Latter, W.~B. \& Black, J.~H. 1991, ApJ, 372, 161

\bibitem[{{Lepp} \& {Shull}(1983)}]{ls83}
{Lepp}, S. \& {Shull}, J.~M. 1983, \apj, 270, 578

\bibitem[{Lepp \& Shull(1984)}]{lepp1984molecules}
Lepp, S. \& Shull, J.~M. 1984, \apj, 280, 465

\bibitem[{Lepp {et~al.}(2002)Lepp, Stancil, \& Dalgarno}]{Lepp_2002}
Lepp, S., Stancil, P.~C., \& Dalgarno, A. 2002, J. Phys. B: Atom., Mol. Opt.
  Phys., 35, R57

\bibitem[{{Linder} {et~al.}(1995){Linder}, {Janev}, \& {Botero}}]{ljb95}
{Linder}, F., {Janev}, R.~K., \& {Botero}, J. 1995, in Atomic and Molecular
  Processes in Fusion Edge Plasmas, ed. R.~K. {Janev}, 397

\bibitem[{Lindinger {et~al.}(1982)Lindinger, Howorka, Maerk, \& Egger}]{ger82}
Lindinger, W., Howorka, F., Maerk, T., \& Egger, F. 1982, in Symposium on
  Atomic and Surface Phys., 1982 Contrib.

\bibitem[{{Lique}(2015)}]{Lique:2015}
{Lique}, F. 2015, \mnras, 453, 810

\bibitem[{Longo {et~al.}(2011)Longo, Coppola, Galli, Palla, \&
  Capitelli}]{bk978-0-7503-1425-1ch1bib61}
Longo, S., Coppola, C.~M., Galli, D., Palla, F., \& Capitelli, M. 2011,
  Rendiconti Lincei, 22, 119

\bibitem[{{Mac Low} \& {Shull}(1986)}]{MAC86}
{Mac Low}, M.~M. \& {Shull}, J.~M. 1986, \apj, 302, 585

\bibitem[{Maoli {et~al.}(1996)Maoli, Ferrucci, Melchiorri, Signore, \&
  Tosti}]{Maoli1996}
Maoli, R., Ferrucci, V., Melchiorri, F., Signore, M., \& Tosti, D. 1996, \apj,
  457, 1

\bibitem[{Maoli {et~al.}(1994)Maoli, Melchiorri, \& Tosti}]{Maoli1994}
Maoli, R., Melchiorri, F., \& Tosti, D. 1994, \apj, 425, 372

\bibitem[{{Martin} {et~al.}(1998){Martin}, {Keogh}, \& {Mandy}}]{MAR98}
{Martin}, P.~G., {Keogh}, W.~J., \& {Mandy}, M.~E. 1998, \apj, 499, 793

\bibitem[{{M{\'e}ndez} {et~al.}(2006){M{\'e}ndez}, {Gordillo-V{\'a}zquez},
  {Herrero}, \& {Tanarro}}]{Mendez2006}
{M{\'e}ndez}, I., {Gordillo-V{\'a}zquez}, F.~J., {Herrero}, V.~J., \&
  {Tanarro}, I. 2006, JPC A, 110, 6060

\bibitem[{{Mielke} {et~al.}(2003){Mielke}, {Peterson}, {Schwenke}, {Garrett},
  {Truhlar}, {Michael}, {Su}, \& {Sutherland}}]{mie03}
{Mielke}, S.~L., {Peterson}, K.~A., {Schwenke}, D.~W., {et~al.} 2003, \prl, 91,
  063201

\bibitem[{{Novosyadlyj} {et~al.}(2022){Novosyadlyj}, {Kulinich}, {Melekh}, \&
  {Shulga}}]{Novosyadlyj2022}
{Novosyadlyj}, B., {Kulinich}, Y., {Melekh}, B., \& {Shulga}, V. 2022, \aap,
  663, A120

\bibitem[{{Novosyadlyj} {et~al.}(2023){Novosyadlyj}, {Kulinich}, {Milinevsky},
  \& {Shulga}}]{Novosyadlyj:2023}
{Novosyadlyj}, B., {Kulinich}, Y., {Milinevsky}, G., \& {Shulga}, V. 2023,
  \mnras, 526, 2724

\bibitem[{{Novosyadlyj} {et~al.}(2018){Novosyadlyj}, {Shulga}, {Han},
  {Kulinich}, \& {Tsizh}}]{Novosyadlyj2018}
{Novosyadlyj}, B., {Shulga}, V., {Han}, W., {Kulinich}, Y., \& {Tsizh}, M.
  2018, \apj, 865, 38

\bibitem[{{Peart} \& {Hayton}(1994)}]{ph94}
{Peart}, B. \& {Hayton}, D.~A. 1994, J. Phys. B Atom. Mol. Phys., 27, 2551

\bibitem[{{Persson, C. M.} {et~al.}(2010){Persson, C. M.}, {Maoli, R.},
  {Encrenaz, P.}, {Hjalmarson, Å.}, {Olberg, M.}, {Rydbeck, G.}, {Signore,
  M.}, {Frisk, U.}, {Sandqvist, Aa.}, \& {Daniel, J. Y.}}]{Persson2010}
{Persson, C. M.}, {Maoli, R.}, {Encrenaz, P.}, {et~al.} 2010, A\&A, 515, A72

\bibitem[{{Planck Collaboration} {et~al.}(2020){Planck Collaboration},
  {Aghanim, N.}, {Akrami, Y.}, {Ashdown, M.}, {Aumont, J.}, {Baccigalupi, C.},
  {Ballardini, M.}, {Banday, A. J.}, {Barreiro, R. B.}, {Bartolo, N.}, {Basak,
  S.}, {Battye, R.}, {Benabed, K.}, {Bernard, J.-P.}, {Bersanelli, M.},
  {Bielewicz, P.}, {Bock, J. J.}, {Bond, J. R.}, {Borrill, J.}, {Bouchet, F.
  R.}, {Boulanger, F.}, {Bucher, M.}, {Burigana, C.}, {Butler, R. C.},
  {Calabrese, E.}, {Cardoso, J.-F.}, {Carron, J.}, {Challinor, A.}, {Chiang, H.
  C.}, {Chluba, J.}, {Colombo, L. P. L.}, {Combet, C.}, {Contreras, D.},
  {Crill, B. P.}, {Cuttaia, F.}, {de Bernardis, P.}, {de Zotti, G.},
  {Delabrouille, J.}, {Delouis, J.-M.}, {Di Valentino, E.}, {Diego, J. M.},
  {Doré, O.}, {Douspis, M.}, {Ducout, A.}, {Dupac, X.}, {Dusini, S.},
  {Efstathiou, G.}, {Elsner, F.}, {Enßlin, T. A.}, {Eriksen, H. K.}, {Fantaye,
  Y.}, {Farhang, M.}, {Fergusson, J.}, {Fernandez-Cobos, R.}, {Finelli, F.},
  {Forastieri, F.}, {Frailis, M.}, {Fraisse, A. A.}, {Franceschi, E.}, {Frolov,
  A.}, {Galeotta, S.}, {Galli, S.}, {Ganga, K.}, {Génova-Santos, R. T.},
  {Gerbino, M.}, {Ghosh, T.}, {González-Nuevo, J.}, {Górski, K. M.},
  {Gratton, S.}, {Gruppuso, A.}, {Gudmundsson, J. E.}, {Hamann, J.}, {Handley,
  W.}, {Hansen, F. K.}, {Herranz, D.}, {Hildebrandt, S. R.}, {Hivon, E.},
  {Huang, Z.}, {Jaffe, A. H.}, {Jones, W. C.}, {Karakci, A.}, {Keihänen, E.},
  {Keskitalo, R.}, {Kiiveri, K.}, {Kim, J.}, {Kisner, T. S.}, {Knox, L.},
  {Krachmalnicoff, N.}, {Kunz, M.}, {Kurki-Suonio, H.}, {Lagache, G.},
  {Lamarre, J.-M.}, {Lasenby, A.}, {Lattanzi, M.}, {Lawrence, C. R.}, {Le
  Jeune, M.}, {Lemos, P.}, {Lesgourgues, J.}, {Levrier, F.}, {Lewis, A.},
  {Liguori, M.}, {Lilje, P. B.}, {Lilley, M.}, {Lindholm, V.}, {López-Caniego,
  M.}, {Lubin, P. M.}, {Ma, Y.-Z.}, {Macías-Pérez, J. F.}, {Maggio, G.},
  {Maino, D.}, {Mandolesi, N.}, {Mangilli, A.}, {Marcos-Caballero, A.}, {Maris,
  M.}, {Martin, P. G.}, {Martinelli, M.}, {Martínez-González, E.},
  {Matarrese, S.}, {Mauri, N.}, {McEwen, J. D.}, {Meinhold, P. R.},
  {Melchiorri, A.}, {Mennella, A.}, {Migliaccio, M.}, {Millea, M.}, {Mitra,
  S.}, {Miville-Deschênes, M.-A.}, {Molinari, D.}, {Montier, L.}, {Morgante,
  G.}, {Moss, A.}, {Natoli, P.}, {Nørgaard-Nielsen, H. U.}, {Pagano, L.},
  {Paoletti, D.}, {Partridge, B.}, {Patanchon, G.}, {Peiris, H. V.}, {Perrotta,
  F.}, {Pettorino, V.}, {Piacentini, F.}, {Polastri, L.}, {Polenta, G.},
  {Puget, J.-L.}, {Rachen, J. P.}, {Reinecke, M.}, {Remazeilles, M.}, {Renzi,
  A.}, {Rocha, G.}, {Rosset, C.}, {Roudier, G.}, {Rubiño-Martín, J. A.},
  {Ruiz-Granados, B.}, {Salvati, L.}, {Sandri, M.}, {Savelainen, M.}, {Scott,
  D.}, {Shellard, E. P. S.}, {Sirignano, C.}, {Sirri, G.}, {Spencer, L. D.},
  {Sunyaev, R.}, {Suur-Uski, A.-S.}, {Tauber, J. A.}, {Tavagnacco, D.}, {Tenti,
  M.}, {Toffolatti, L.}, {Tomasi, M.}, {Trombetti, T.}, {Valenziano, L.},
  {Valiviita, J.}, {Van Tent, B.}, {Vibert, L.}, {Vielva, P.}, {Villa, F.},
  {Vittorio, N.}, {Wandelt, B. D.}, {Wehus, I. K.}, {White, M.}, {White, S. D.
  M.}, {Zacchei, A.}, \& {Zonca, A.}}]{Planck2020}
{Planck Collaboration}, {Aghanim, N.}, {Akrami, Y.}, {et~al.} 2020, A\&A, 641,
  A6

\bibitem[{{Poulaert} {et~al.}(1978){Poulaert}, {Brouillard}, {Claeys},
  {McGowan}, \& {Van Wassenhove}}]{POU78}
{Poulaert}, G., {Brouillard}, F., {Claeys}, W., {McGowan}, J.~W., \& {Van
  Wassenhove}, G. 1978, J. Phys. B Atom. Mol. Phys., 11, L671

\bibitem[{{Puy} {et~al.}(1993){Puy}, {Alecian}, {Le Bourlot}, {Leorat}, \&
  {Pineau Des Forets}}]{Puy1993}
{Puy}, D., {Alecian}, G., {Le Bourlot}, J., {Leorat}, J., \& {Pineau Des
  Forets}, G. 1993, \aap, 267, 337

\bibitem[{Puy \& Signore(2007)}]{puy2007primordial}
Puy, D. \& Signore, M. 2007, New Astron. Rev., 51, 411

\bibitem[{{Ramaker} \& {Peek}(1976)}]{RAM76}
{Ramaker}, D.~E. \& {Peek}, J.~M. 1976, \pra, 13, 58

\bibitem[{{Roueff} {et~al.}(2019){Roueff}, {Abgrall}, {Czachorowski},
  {Pachucki}, {Puchalski}, \& {Komasa}}]{Roueff:2019}
{Roueff}, E., {Abgrall}, H., {Czachorowski}, P., {et~al.} 2019, \aap, 630, A58

\bibitem[{{Savin}(2002)}]{sav02}
{Savin}, D.~W. 2002, \apj, 566, 599

\bibitem[{{Savin} {et~al.}(2004){Savin}, {Krsti{\'c}}, {Haiman}, \&
  {Stancil}}]{SAV04}
{Savin}, D.~W., {Krsti{\'c}}, P.~S., {Haiman}, Z., \& {Stancil}, P.~C. 2004,
  \apjl, 607, L147

\bibitem[{{Sbordone} {et~al.}(2010){Sbordone}, {Bonifacio}, {Caffau}, {Ludwig},
  {Behara}, {Gonz{\'a}lez Hern{\'a}ndez}, {Steffen}, {Cayrel}, {Freytag},
  {van't Veer}, {Molaro}, {Plez}, {Sivarani}, {Spite}, {Spite}, {Beers},
  {Christlieb}, {Fran{\c{c}}ois}, \& {Hill}}]{Sbo10}
{Sbordone}, L., {Bonifacio}, P., {Caffau}, E., {et~al.} 2010, \aap, 522, A26

\bibitem[{{Schauer} {et~al.}(1989){Schauer}, {Jefferts}, {Barlow}, \&
  {Dunn}}]{Schauer1989}
{Schauer}, M.~M., {Jefferts}, S.~R., {Barlow}, S.~E., \& {Dunn}, G.~H. 1989,
  \jcp, 91, 4593

\bibitem[{{Schleicher} {et~al.}(2008){Schleicher}, {Galli}, {Palla},
  {Camenzind}, {Klessen}, {Bartelmann}, \& {Glover}}]{Schleicher:2008}
{Schleicher}, D.~R.~G., {Galli}, D., {Palla}, F., {et~al.} 2008, \aap, 490, 521

\bibitem[{{Schulz} \& {Asundi}(1967)}]{sa67}
{Schulz}, G.~J. \& {Asundi}, R.~K. 1967, Phys. Rev., 158, 25

\bibitem[{{Seager} {et~al.}(1999){Seager}, {Sasselov}, \& {Scott}}]{Seager1999}
{Seager}, S., {Sasselov}, D.~D., \& {Scott}, D. 1999, \apjl, 523, L1

\bibitem[{Sethi {et~al.}(2008)Sethi, Nath, \&
  Subramanian}]{sethi2008primordial}
Sethi, S.~K., Nath, B.~B., \& Subramanian, K. 2008, \mnras, 387, 1589

\bibitem[{{Shapiro} \& {Kang}(1987)}]{sk87}
{Shapiro}, P.~R. \& {Kang}, H. 1987, \apj, 318, 32

\bibitem[{{Shavitt}(1959)}]{s59}
{Shavitt}, I. 1959, \jcp, 31, 1359

\bibitem[{Signore \& Puy(2009)}]{signore2009cosmic}
Signore, M. \& Puy, D. 2009, EPJ C, 59, 117

\bibitem[{{Soon}(1992)}]{Soon1992}
{Soon}, W.~H. 1992, \apj, 394, 717

\bibitem[{Stancil {et~al.}(1996)Stancil, Lepp, \&
  Dalgarno}]{bk978-0-7503-1425-1ch1bib90}
Stancil, P.~C., Lepp, S., \& Dalgarno, A. 1996, ApJ, 458, 401

\bibitem[{{Stancil} {et~al.}(1998){Stancil}, {Lepp}, \&
  {Dalgarno}}]{Stancil1998}
{Stancil}, P.~C., {Lepp}, S., \& {Dalgarno}, A. 1998, \apj, 509, 1

\bibitem[{{Str{\"o}mholm} {et~al.}(1995){Str{\"o}mholm}, {Schneider},
  {Sundstr{\"o}m}, {Carata}, {Danared}, {Datz}, {Dulieu}, {K{\"a}llberg}, {Af
  Ugglas}, {Urbain}, {Zengin}, {Suzor-Weiner}, \& {Larsson}}]{sss95}
{Str{\"o}mholm}, C., {Schneider}, I.~F., {Sundstr{\"o}m}, G., {et~al.} 1995,
  \pra, 52, R4320

\bibitem[{{Theard} \& {Huntress}(1974)}]{Theard1974}
{Theard}, L.~P. \& {Huntress}, W.~T. 1974, \jcp, 60, 2840

\bibitem[{{Trevisan} \& {Tennyson}(2002{\natexlab{a}})}]{tt02a}
{Trevisan}, C.~S. \& {Tennyson}, J. 2002{\natexlab{a}}, Plasma Phys. and
  Control. Fusion, 44, 1263

\bibitem[{{Trevisan} \& {Tennyson}(2002{\natexlab{b}})}]{tt02b}
{Trevisan}, C.~S. \& {Tennyson}, J. 2002{\natexlab{b}}, Plasma Phys. and
  Control. Fusion, 44, 2217

\bibitem[{{Vonlanthen} {et~al.}(2009){Vonlanthen}, {Rauscher}, {Winteler},
  {Puy}, {Signore}, \& {Dubrovich}}]{Vonlanthen2009}
{Vonlanthen}, P., {Rauscher}, T., {Winteler}, C., {et~al.} 2009, \aap, 503, 47

\bibitem[{Walkauskas \& Kaufman(1975)}]{wk75}
Walkauskas, L. \& Kaufman, F. 1975, Symposium (International) on Combustion,
  15, 691, fifteenth Symposium (International) on Combustion

\bibitem[{{Walmsley} {et~al.}(2004){Walmsley}, {Flower}, \& {Pineau des
  For{\^e}ts}}]{wfp04}
{Walmsley}, C.~M., {Flower}, D.~R., \& {Pineau des For{\^e}ts}, G. 2004, \aap,
  418, 1035

\bibitem[{{Wang} \& {Stancil}(2002)}]{ws02}
{Wang}, J.~G. \& {Stancil}, P.~C. 2002, Phys. Scr., T96, 72

\bibitem[{{Wishart}(1979)}]{WIS79}
{Wishart}, A.~W. 1979, \mnras, 187, 59P

\bibitem[{{Xu} \& {Fabrikant}(2001)}]{xf01}
{Xu}, Y. \& {Fabrikant}, I.~I. 2001, APL, 78, 2598

\bibitem[{{Zygelman} {et~al.}(1989){Zygelman}, {Dalgarno}, {Kimura}, \&
  {Lane}}]{z89}
{Zygelman}, B., {Dalgarno}, A., {Kimura}, M., \& {Lane}, N.~F. 1989, \pra, 40,
  2340

\end{thebibliography}

\onecolumn

\begin{appendix}

\section{Absorption signals of H$_2$, HD, and HeH$^+$}

\begin{figure}[H]
\begin{center}
   \includegraphics[width=0.7\textwidth]{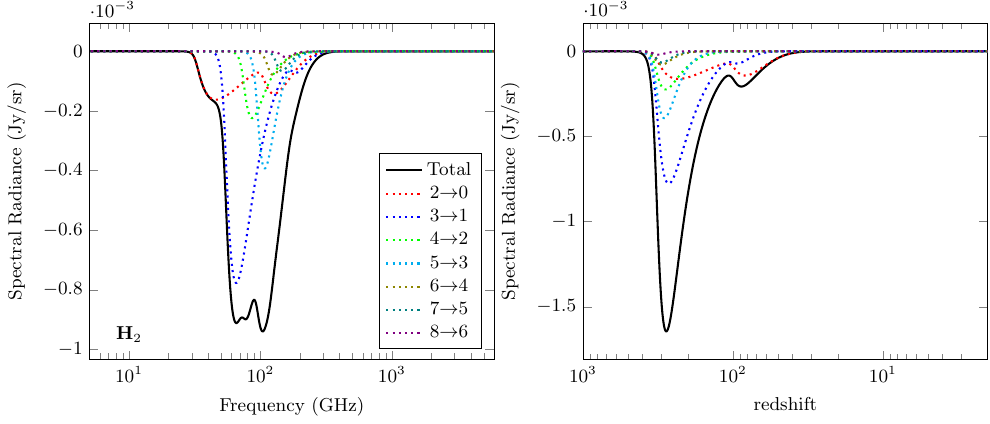}
      \includegraphics[width=0.7\textwidth]{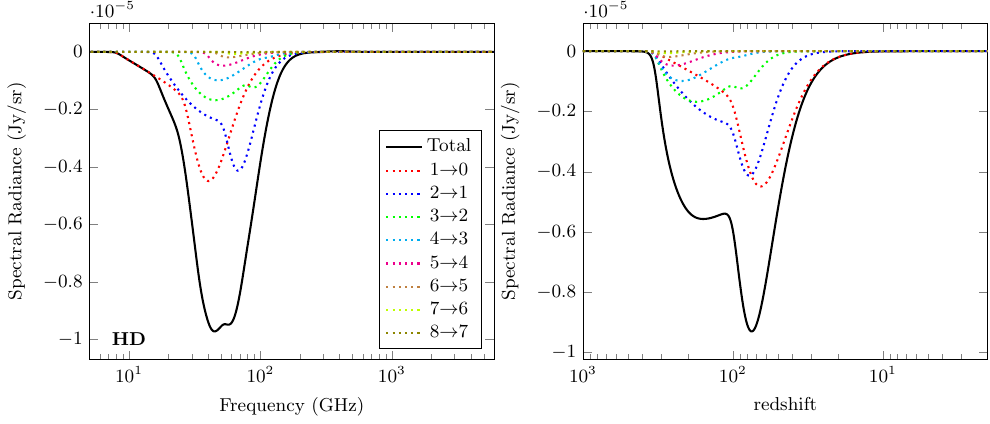}
         \includegraphics[width=0.7\textwidth]{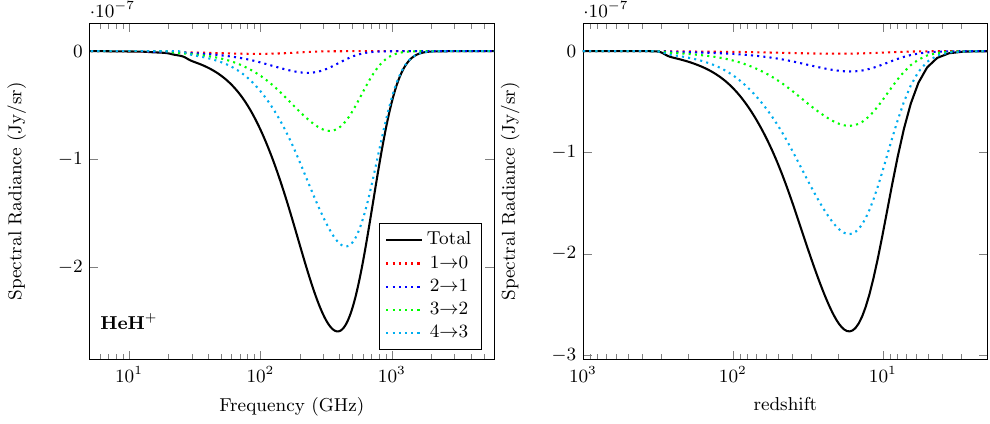}
\end{center}
\caption{ Left panels: absorption signal in the rotational lines of
  molecules H$_2$ (upper row), HD (middle row), and HeH$^+$ (bottom
  row) -- doted coloured lines show contributions from separate
  transitions, and the black line shows their superposition. Right
  panels: redshifts at which absorption occurred for corresponding
  transitions.}
\label{mol_sygnals}
\end{figure}

\newpage

\section{Detailed description of the chemical model}

\small\tiny
\begin{longtable}{|p{0.05\textwidth}|p{0.22\textwidth}|p{0.5\textwidth}|p{0.1\textwidth}|p{0.03\textwidth}|}
  \caption{List of reactions included in our chemical model.} \label{tab:long} \\
\hline \multicolumn{1}{|p{0.05\textwidth}|}{\textbf{Tag}} & \multicolumn{1}{|p{0.22\textwidth}|}{\textbf{Reaction}} & \multicolumn{1}{p{0.5\textwidth}|}{\textbf{Rate coefficient $({\rm cm}^{3} \: {\rm s}^{-1},\, {\rm cm}^{6} \: {\rm s}^{-1},\, {\rm s}^{-1})$}} & \multicolumn{1}{|p{0.1\textwidth}|}{} & \multicolumn{1}{p{0.03\textwidth}|}{\textbf{Ref.}} \\ \hline
\endfirsthead

\multicolumn{5}{c}
{{\bfseries \tablename\ \thetable{} -- continued from previous page}} \\

\hline \multicolumn{1}{|p{0.05\textwidth}|}{\textbf{Tag}} & \multicolumn{1}{|p{0.22\textwidth}|}{\textbf{Reaction}} & \multicolumn{1}{p{0.5\textwidth}|}{\textbf{Rate coefficient $({\rm cm}^{3} \: {\rm s}^{-1},\, {\rm cm}^{6} \: {\rm s}^{-1},\, {\rm s}^{-1})$}} & \multicolumn{1}{|p{0.1\textwidth}|}{} & \multicolumn{1}{p{0.03\textwidth}|}{\textbf{Ref.}} \\ \hline
\endhead

\hline \multicolumn{5}{|r|}{{Continued on next page}} \\ \hline
\endfoot

\hline \hline
\endlastfoot
\hypertarget{H_01}{H\_01}  & $\bm{\Hp + \me  \rightarrow  \mH + \gamma}$ & $k_{rec} = C_H\cdot \alpha_H;$ &  &  1 \\
& & $\phantom{k_{rec} = } \mbox{} \alpha_H =  F\cdot a_1\cdot 10^{-19}t^{a_2}/(1.0+a_3t^{a_4}),$ & & \\
& & $\phantom{k_{rec} = } \mbox{} F=1.14,\, a_1 = 4.309,\, a_2 = -0.6166, $ & & \\
& & $\phantom{k_{rec} = } \mbox{} a_3=0.6703,\, a_4=0.5300,\, t=T/10^4$ & & \\
& & $\phantom{k_{rec} = } \mbox{} C_H = {(1+K_HA_{2s1sH}n_{HI}) \over (1+K_H(A_{2s1sH} + \beta_H)n_{HI})}$ & & \\
& & $\phantom{k_{rec} = } \mbox{} \beta_{\rm H} = \alpha_{\rm H}\cdot (2\pi m_{\rm e} k_B T_{\rm R}/h^2)^{3/2} \exp(-E_{2s}/kT_{\rm R})$  & & \\
& & $\phantom{k_{rec} = } \mbox{} K_{\rm H}\equiv\lambda_{\rm Ly\alpha}^3/(8\pi H(z)),\, \lambda_{\rm Ly\alpha} = 121.5682\,\mathrm{nm}$ & & \\
& & $\phantom{k_{rec} = } \mbox{} E_{\rm 2s} = 5.446756\times 10^{-19}\,\mathrm{J}, \, \Lambda_{\rm H} = 8.22458\,{\rm s}^{-1}$ & & \\
\hypertarget{H_02}{H\_02}  & $\bm{\mH + \gamma  \rightarrow  \Hp + \me }$ & $k_{ion} = C_H\cdot \beta_H \cdot\exp(-E_{2s1s}/kT_{\rm M});$ &  &  1 \\
& & $\phantom{k_{rec} = } \mbox{} E_{\rm 2s1s} = 1.634033\times 10^{-18}\,\mathrm{J}$ & & \\

\hypertarget{H_03}{H\_03}  &  $\bm{\mH + \me  \rightarrow  \Hm + \gamma }$ & $ {\rm dex}[-17.845 + 0.762 \log{T}$ & $T \le 6000 \: {\rm K}$ & 2 \\
& & $\phantom{ {\rm dex}[} \mbox{}+ 0.1523 (\log{T})^{2}$ & & \\
& & $\phantom{ {\rm dex}[} \mbox{}- 0.03274 (\log{T})^{3}] $ & & \\
& & $ \phantom{}  {\rm dex}[-16.4199 + 0.1998 (\log{T})^{2}$ & $T > 6000 \: {\rm K}$ & \\
& & $ \phantom{ {\rm dex}[} \mbox{}-5.447  \times 10^{-3}  (\log{T})^{4}$ & & \\
& & $ \phantom{ {\rm dex}[} \mbox{}+ 4.0415 \times 10^{-5} (\log{T})^{6}]$  & & \\

\hypertarget{H_04}{H\_04} & $\bm{\Hm + \gamma \to \mH + \me}$ & $0.11T_{\rm R}^{2.13}\exp\left(-8823.0/T_{\rm R}\right)$ &  & 3 \\

\hypertarget{H_05}{H\_05} & $\bm{\Hm  + \mH  \rightarrow \mHt + \me}$ & $1.5\times 10^{-9}$ & $T<300.0$ & 3 \\
& & $4.0\times 10^{-9}T^{-0.17}$ &  $T \ge 300.0$ & \\
\hypertarget{H_06}{H\_06} & $\bm{\Hm  + \Hp  \rightarrow \mH + \mH}$ & $2.4\times 10^{-6}(1.0+T/20000)/\sqrt{T}$  & & 4 \\
\hypertarget{H_07}{H\_07}  &  $\bm{\mH + \Hp  \to  \mHtp + \gamma }$ &
$ {\rm dex}[-19.38 - 1.523 \log{T} $ & & 5 \\
& & $\phantom{{\rm dex}[} \mbox{} + 1.118 (\log{T})^{2}  - 0.1269 (\log{T})^{3}]$ & & \\

\hypertarget{H_08}{H\_08} & $\bm{\mHtp + \gamma \to \mH + \Hp}$ & $20.0T_{\rm R}^{1.59}\exp\left(-82000/T_{\rm R}\right)$  & $v=0$ & 3 \\
& & $1.63\times 10^7\exp\left(-32400.0/T_{\rm R}\right)$ & LTE &  \\

\hypertarget{H_09}{H\_09} & $\bm{\mH + \mHtp \rightarrow \mHt + \Hp}$ & $6.4 \times 10^{-10}$ & & 6 \\
\hypertarget{H_10}{H\_10}  &  $\bm{\mHt + \Hp  \rightarrow  \mHtp + \mH}$ &
$[- 3.3232183 \times 10^{-7}$ & & 7 \\
& & $\phantom{[}  \mbox{} + 3.3735382 \times 10^{-7}  \log{T}$  & & \\
& & $\phantom{[}  \mbox{} - 1.4491368 \times 10^{-7}  (\log{T})^2$ & & \\
& & $\phantom{[}  \mbox{} + 3.4172805 \times 10^{-8}  (\log{T})^3$ & & \\
& & $\phantom{[}  \mbox{} - 4.7813720 \times 10^{-9}  (\log{T})^4$ & & \\
& & $\phantom{[}  \mbox{} + 3.9731542 \times 10^{-10} (\log{T})^5$ & & \\
& & $\phantom{[}  \mbox{}  - 1.8171411 \times 10^{-11}  (\log{T})^6$ & & \\
& & $\phantom{[}  \mbox{}  + 3.5311932 \times 10^{-13} (\log{T})^7 ]$ & & \\
& & $\phantom{[} \mbox{} \times \exp \left(\frac{-21237.15}{T} \right)$ & & \\
\hypertarget{H_11}{H\_11}  &  $\bm{\Hp + \Hm  \rightarrow  \mHtp + \me}$ &
$  6.9 \times 10^{-9}  T^{-0.35}$  & $T \le 8000 \: {\rm K}$ & 8 \\
& & $\phantom{} 9.6 \times 10^{-7} T^{-0.90}$ & $T > 8000 \: {\rm K}$ & \\
\hypertarget{H_12}{H\_12}  &  $\bm{\mHtp + \me  \rightarrow  \mH + \mH}$ &
$4.2278 \times 10^{-14} - 2.3088\times 10^{-17} T + 7.3428\times 10^{-21} T^2$ & & \\
& & $\phantom{[} - 7.5474\times 10^{-25} T^3 + 3.3468\times 10^{-29} T^4 - 5.528\times 10^{-34} T^5$ & & 9 \\

\hypertarget{H_13}{H\_13} & $\bm{\mHtp + \mHt \to \htp + \mH}$ & $2.0\times 10^{-9}$ &  & 10 \\

\hypertarget{H_14}{H\_14} & $\bm{\mHt + \gamma \to \mHtp + \me}$ & $290.0T_{\rm R}^{1.56}\exp\left(-178500.0/T_{\rm R}\right )$ &  & 3 \\
\hypertarget{H_15}{H\_15} & $\bm{\mHt + \gamma \to \mH + \mH}$ & $2394.9T_{\rm R}^{0.62681}\left(1.0+2.4635T_{\rm R}^{0.56957}\right)/\exp\left(140050/T_{\rm R}\right)$ & $T_{\rm R}<15000$ & 11 \\
 &  & $6.4471\times 10^7 T_{\rm R}^{0.322}\left(1.0+6.5341\times 10^{-8}T_{\rm R}^{1.4752}\right)\exp\left(-\frac{153570}{T_{\rm R}}\right)$ &  $T_{\rm R}\ge 15000$  &  \\
\hypertarget{H_16}{H\_16} & $\bm{\htp + \mH \to \mHtp + \mHt}$ & $7.7\times 10^{-9}\exp\left(-17560.0/T_{\rm M}\right)$ &  & 3 \\
\hypertarget{H_17}{H\_17} & $\bm{\htp + \me \to \mHt + \mH}$ & $4.6\times 10^{-6}T_{\rm M}^{-0.65}$ &  & 3 \\
\hypertarget{H_18}{H\_18} & $\bm{\mHt + \Hp \to \htp + \gamma}$ & $10^{-16}$ &  & 3 \\

\hypertarget{H_19}{H\_19} & $\bm{\mHt + \mH  \rightarrow  \mH + \mH + \mH}$  &
$6.67 \times 10^{-12} T^{0.5} \exp \left[-(1+ \frac{63593}{T}) \right]$ & v=0 & 12 \\
& & $\phantom{}  3.52 \times 10^{-9} \expf{-}{43900}{T}$ & LTE & 13 \\
\hypertarget{H_20}{H\_20} &  $\bm{\mH + \me  \rightarrow  \Hp + \me + \me}$ &
$\exp[-3.271396786 \times 10^{1}$ & & 14 \\
& & $\phantom{\exp[} \mbox{}  + 1.35365560 \times 10^{1} \ln T_{\rm eV}$ & & \\
& & $\phantom{\exp[} \mbox{}  - 5.73932875 \times 10^{0} (\ln T_{\rm eV})^{2}$  & & \\
& & $\phantom{\exp[} \mbox{}  + 1.56315498 \times 10^{0} (\ln T_{\rm eV})^{3}$ & & \\
& & $\phantom{\exp[} \mbox{}  -  2.87705600 \times 10^{-1} (\ln T_{\rm eV})^{4}$ & & \\
& & $\phantom{\exp[} \mbox{}  + 3.48255977 \times 10^{-2} (\ln T_{\rm eV})^{5}$ & & \\
& & $\phantom{\exp[} \mbox{}   - 2.63197617 \times 10^{-3} (\ln T_{\rm eV})^{6}$ & & \\
& & $\phantom{\exp[} \mbox{}  + 1.11954395\times 10^{-4} (\ln T_{\rm eV})^{7}$ & & \\
& & $\phantom{\exp[} \mbox{}   -  2.03914985 \times 10^{-6} (\ln T_{\rm eV})^{8}]$ & & \\

\hypertarget{H_21}{H\_21} & $\bm{\htp + \gamma \rightarrow \mHt + \Hp}$ & $4.5\times 10^{8}\exp\left(-\frac{232258}{T_{\rm R}}\right)$ & & 15 \\

\hypertarget{H_22}{H\_22} & $\bm{\htp + \me \rightarrow 3\mH}$ & the same as for \hyperlink{H_17}{\color{blue}H\_17} & & \\

\hypertarget{H_23}{H\_23} & $\bm{\mH + \me \rightarrow \Hp + 2\me}$ &
$6.5023\times 10^{-9}T_{\rm eV}^{0.48931}\exp\left(-\frac{12.89365}{T_{\rm eV}}\right)$ & & 16 \\
\hypertarget{H_24}{H\_24} & $\bm{\Hp + \mH + \mH \rightarrow \Hp + \mHt}$ & $1.145\times 10^{-29}T^{-1.12}$  &  & 17 \\

\hypertarget{H_25}{H\_25} & $\mH + \mH + \mH \rightarrow \mHt + \mH$ & $1.14\times 10^{-31}T^{-0.38}$  & $T<300$~K & 18 \\
&  & $3.9\times 10^{-30}T^{-1}$  & $T\ge 300$~K &  \\

\hypertarget{H_26}{H\_26} & $\mHtp + \gamma \to 2\Hp + \me$ & $90.0 T_{\rm R}^{1.48}\exp\left(-335000.0/T_{\rm R}\right)$ &  & 3 \\
\hypertarget{H_27}{H\_27} & $\mHt + \me \rightarrow \Hm + \mH$ &
$2.7 \times 10^{-8} T^{-1.27} \expf{-}{43000}{T}$ & & 19 \\

\hypertarget{H_28}{H\_28} & $\mHt + \me  \rightarrow  \mH + \mH +  \me$ &
$4.49 \times 10^{-9} T^{0.11} \expf{-}{101858}{T}$ & $v=0$ & 20 \\
& & $\phantom{} 1.91 \times 10^{-9} T^{0.136} \expf{-}{53407.1}{T}$ & LTE & 20 \\

\hypertarget{H_29}{H\_29} & $\mHt + \mHt \rightarrow  \mHt + \mH + \mH$ &
$\frac{5.996 \times 10^{-30} T^{4.1881}}{(1.0 + 6.761 \times 10^{-6} T)^{5.6881}}
\exp \left(-\frac{54657.4}{T} \right)$ & $v=0$ & 21 \\
& & $\phantom{} 1.3 \times 10^{-9} \expf{-}{53300}{T}$ & LTE & 22 \\
\hypertarget{H_30}{H\_30} & $\mHt + \He \rightarrow \mH + \mH + \He$ &
${\rm dex} \left[ -27.029 + 3.801 \log{T} - \frac{29487}{T} \right]$ & $v=0$ & 23 \\
& & $\phantom{}  {\rm dex} \left[ -2.729 -1.75 \log{T} - \frac{23474}{T} \right]$ & LTE & 23 \\

\hypertarget{H_31}{H\_31}  &  $\Hm + \me  \rightarrow  \mH + \me + \me $ &
$ \exp [-1.801849334 \times 10^{1}$ & & 14 \\
& & $\phantom{\exp [} \mbox{} + 2.36085220 \times 10^{0} \ln T_{\rm eV}$ & &  \\
& & $\phantom{\exp [} \mbox{} - 2.82744300 \times 10^{-1} (\ln T_{\rm eV})^{2}$ & & \\
& & $\phantom{\exp [} \mbox{}  +1.62331664\times 10^{-2} (\ln T_{\rm eV})^{3}$ & & \\
& & $\phantom{\exp [} \mbox{} -3.36501203 \times 10^{-2} (\ln T_{\rm eV})^{4}$ & & \\
& & $\phantom{\exp [} \mbox{}   +1.17832978\times 10^{-2} (\ln T_{\rm eV})^{5}$ & & \\
& & $\phantom{\exp [} \mbox{}  -1.65619470\times 10^{-3} (\ln T_{\rm eV})^{6}$ & & \\
& & $\phantom{\exp [} \mbox{}   +1.06827520\times 10^{-4} (\ln T_{\rm eV})^{7}$ & & \\
& & $\phantom{\exp [} \mbox{}  -2.63128581\times 10^{-6} (\ln T_{\rm eV})^{8} ]$ & & \\
\hypertarget{H_32}{H\_32}  &  $\Hm + \mH  \rightarrow  \mH + \mH + \me $ &
$2.5634 \times 10^{-9} T_{\rm eV}^{1.78186}$ & $T_{\rm eV} \leq 0.1 \: \rm{eV}$ & 14 \\
& & $\exp[-2.0372609 \times 10^{1}$ & $T_{\rm eV} > 0.1 \: \rm{eV}$  & \\
& & $\phantom{\exp [} \mbox{}+1.13944933 \times 10^{0} \ln T_{\rm eV}$ & & \\
& & $\phantom{\exp [} \mbox{}-1.4210135 \times 10^{-1} (\ln T_{\rm eV})^{2}$ & & \\
& & $\phantom{\exp [} \mbox{}+8.4644554 \times 10^{-3} (\ln T_{\rm eV})^{3}$ & & \\
& & $\phantom{\exp [} \mbox{}-1.4327641 \times 10^{-3} (\ln T_{\rm eV})^{4}$  & & \\
& & $\phantom{\exp [} \mbox{}+2.0122503 \times 10^{-4} (\ln T_{\rm eV})^{5}$ & & \\
& & $\phantom{\exp [} \mbox{}+8.6639632 \times 10^{-5} (\ln T_{\rm eV})^{6}$ & & \\
& & $\phantom{\exp [} \mbox{}-2.5850097 \times 10^{-5} (\ln T_{\rm eV})^{7}$ & & \\
& & $\phantom{\exp [} \mbox{}+ 2.4555012\times 10^{-6} (\ln T_{\rm eV})^{8}$ & & \\
& & $\phantom{\exp [} \mbox{} -8.0683825\times 10^{-8} (\ln T_{\rm eV})^{9}]$ & & \\

\hypertarget{H_33}{H\_33} & $\Hm + \mHtp \rightarrow \mHt + \mH$ &
$1.4 \times 10^{-7} \left(\frac{T}{300}\right)^{-0.5}$  & & 24 \\
\hypertarget{H_34}{H\_34} & $\Hm + \mHtp \rightarrow \mH + \mH + \mH$ &
$1.4 \times 10^{-7} \left(\frac{T}{300}\right)^{-0.5}$  & & 24 \\

\hypertarget{H_35}{H\_35} & $\mH + \mH + \mHt \rightarrow \mHt + \mHt$ & $ 0.143\times 10^{-31}T^{-0.38}$  & $T<300$~K & 18 \\
&  & $0.486\times 10^{-30}T^{-1}$  & $T\ge 300$~K & 18 \\

\hypertarget{H_36}{H\_36} & $\mH + \mH + \He \rightarrow \mHt + \He$ & $ 6.9 \times 10^{-32} T^{-0.4}$ & & 25 \\

\hypertarget{H_37}{H\_37} & $\mHtp + \mH \to 2\mH + \Hp$ & $\exp\left(-32.912 + 6.9498\times 10^{-5}T - 3.3248\times 10^4/T \right. $ &  & 9 \\
&  & $\phantom{\exp(}\left. - 4.08\times 10^{-9}T^2\right)$ &  &  \\
\hypertarget{H_38}{H\_38} & $\mHt + \Hp \to 2\mH + \Hp$ & $\exp\left(-33.404 + 2.0148\times 10^{-4}T - 5.2674\times 10^4/T \right.$ &  & 9 \\
&  & $\phantom{\exp(}\left. - 1.0196\times 10^{-8}T^2\right)$ &  &  \\

\hypertarget{H_39}{H\_39} & $\htp + \mHt \rightarrow \htp + 2\mH$ & $0.3\times 10^{-10}\left(\frac{T}{300}\right)^{0.5}\exp\left(-\frac{52000}{T}\right)$  &  & 26 \\

\hypertarget{H_40}{H\_40} & ${\htp + \me \rightarrow \mHtp + \mH + \me}$ & $4.8462\times 10^{-7}T_{\rm eV}^{-0.04975}\exp\left(-\frac{19.165665}{T_{\rm eV}}\right)$ & & 16 \\

\hypertarget{H_41}{H\_41} & ${\mHtp + \me \rightarrow \Hp + \mH + \me}$ & $1.0702\times 10^{-7}T_{\rm eV}^{0.04876}\exp\left(-\frac{9.69028}{T_{\rm eV}}\right)$ & & 16 \\

 \hypertarget{H_42}{H\_42} & $\mH + \mH \rightarrow \mH + \Hp + \me$ &
$1.45\times 10^{-15}T^{0.5}\left(1.0+\frac{T}{26280}\right)\exp\left(-\frac{78841}{T}\right)$ & & 27 \\

\hypertarget{H_43}{H\_43} & $\Hp + \mH + \mH \rightarrow \htp + \mH$ & $1.238\times 10^{-29}T^{-1.046}$  &  & 17 \\

\hypertarget{H_44}{H\_44} & ${\htp + \Hm \rightarrow \mHt + \mHt}$ & $2.3\times 10^{-7}\left(\frac{T}{300}\right)^{-0.5}$ & & 28 \\

\hline

\hypertarget{He_01}{He\_01}  & $\bm{\Hep + \me  \rightarrow  \He + \gamma}$ & $k_{rec} = C_{\rm HeI}\cdot \alpha_{\rm HeI};$ &  & 1 \\
& & $\phantom{k_{1} = } \mbox{} \alpha_{\rm HeI} =q\left[\sqrt{T_{\rm M}\over T_2}\left(1+\sqrt{T_{\rm  M}\over T_2}\right)^{1-p}
 \left(1+\sqrt{T_{\rm M}\over T_1}\right)^{1+p}\right]^{-1}\! \mathrm{m^{3}s^{-1}},$ & & \\
& & $\phantom{k_{1} = } \mbox{} q=10^{-16.744}$, $p=0.711$, $T_1=10^{5.114}\,$K, $T_2 = 3\,$K.  & &  \\
& & $\phantom{k_{1} = } \mbox{} C_{\rm HeI} = {\left(1 + K_{\rm HeI} \Lambda_{\rm He} n_{\rm HeI} \exp\left(-E_{\rm 2p2s}/k_BT_{\rm M}\right)\right)
\over \left(1+K_{\rm HeI}   (\Lambda_{\rm He} + \beta_{\rm HeI}) n_{\rm HeI} \exp\left(-E_{\rm 2p2s}/kT_{\rm M}\right)\right)},$   & & \\
& & $\phantom{k_{1} = } \mbox{} \beta_{\rm HeI} = \alpha_{\rm HeI}\cdot \frac14(2\pi m_{\rm e} k_B T_{\rm M}/h^2)^{3/2} \exp\left(-E_{2s}/kT_{\rm M}\right)$  & & \\

& & $\phantom{k_{1} = } \mbox{} K_{\rm HeI} \equiv\lambda_{\rm Ly\alpha}^3/(8\pi H(z)),\, \lambda_{\rm Ly\alpha} = 58.4334\,\mathrm{nm},\, \Lambda_{\rm He} = 51.3\,{\rm s}^{-1}.$ & & \\
& & $\phantom{k_{1} = } \mbox{} E_{\rm 2p2s} = 9.64908313\times 10^{-20}\,\mathrm{J},\, E_{\rm 2s} = 6.363254\times 10^{-19}\,\mathrm{J}$ & & \\

\hypertarget{He_02}{He\_02}  & $\bm{\He + \gamma  \rightarrow  \Hep + \me }$ & $k_{ion} = C_{\rm HeI}\cdot \beta_{\rm HeI}\exp(-E_{2s1s}/k_BT_{\rm M});$ &  & 1 \\
& & $\phantom{k_{1} = } \mbox{} E_{\rm 2s1s} = 3.30301387\times 10^{-18}\,\mathrm{J}$ & &  \\

\hypertarget{He_03}{He\_03} & $\bm{\He + \Hp \to \HeHp + \gamma}$ & $1.56\times10^{-20} \left(\frac{T}{300}\right)^{-0.374}\exp\left(-\frac{T}{46000}\right) $ &  & 29 \\
 &  & $\phantom{ = } + 11.8\times10^{-20}\left(\frac{T}{300}\right)^{-0.244}\exp\left(-\frac{T}{2890}\right)$ &  &  \\
\hypertarget{He_04}{He\_04} & $\bm{\HeHp + \mH \to \He + \mHtp}$ & $1.27228\times 10^{-9}\left(1.0+0.904409\frac{207.632}{8.314472T}\right)^{-1.0/0.904409}$ &  & 30 \\
\hypertarget{He_05}{He\_05} & $\bm{\HeHp + \gamma \to \He + \Hp}$ & $2.03097\times 10^8T_{\rm R}^{-1.20281}\exp\left(-\frac{24735}{T_{\rm R}}\right)$ &  &  31 \\

\hypertarget{He_06}{He\_06} & $\bm{\Hepp + \me \rightarrow \Hep + \gamma}$ &
 $k_{{\rm A}} = 2.538 \times 10^{-13} \left(\frac{1262456}{T}\right)^{1.503}$ & Case A & 32 \\
& & $\phantom{k_{{\rm B}} = } \mbox{} \times
 [1.0+ \left(\frac{2418500}{T}\right)^{0.470}]^{-1.923}$ & & \\
& & $k_{{\rm B}} = 5.506 \times 10^{-14} \left(\frac{1262456}{T}\right)^{1.500}$ & Case B & 32 \\
& & $\phantom{k_{{\rm B}} = } \mbox{} \times
\left[1.0+ \left(\frac{460752}{T}\right)^{0.407}\right]^{-2.242} $ & & \\

\hypertarget{He_07}{He\_07} & $\bm{\Hep + \gamma \to \Hepp + \me}$ & $50.0T_{\rm R}^{1.63}\exp\left(-\frac{590000.0}{T_{\rm R}}\right)$ &  & 3 \\

\hypertarget{He_08}{He\_08} & $\bm{\He + \Hp \rightarrow \Hep + \mH}$ &
$1.26 \times 10^{-9} T^{-0.75} \expf{-}{127500}{T}$ &
$ T \leq 10000 \: {\rm K}$ & 33 \\
& & $\phantom{} 4.0 \times 10^{-37} T^{4.74}$ & $T > 10000 \: {\rm K}$ & \\
\hypertarget{He_09}{He\_09} & $\bm{\Hep + \mH \rightarrow \He + \Hp}$ &
$1.2 \times 10^{-15} \left(\frac{T}{300}\right)^{0.25}$ & & 34 \\

\hypertarget{He_10}{He\_10} & $\bm{\He + \mHtp \to \HeHp + \mH}$ & $3.0\times 10^{-10}\exp\left(-\frac{6717.0}{T}\right)$ &  & 35 \\
\hypertarget{He_11}{He\_11} & $\bm{\Hep + \mH \to \HeHp + \gamma}$ & $4.16\times 10^{-16}T^{-0.37}\exp\left(-\frac{T}{87600}\right)$ &  & 59 \\
\hypertarget{He_12}{He\_12} & $\bm{\HeHp + \me \to \He + \mH}$ & $3.0\times 10^{-8}T^{-0.5}$ &  & 28 \\
\hypertarget{He_13}{He\_13} & $\bm{\HeHp + \mHt \to \htp + \He}$ & $1.80\times 10^{-9}$ &  & 28 \\
\hypertarget{He_14}{He\_14} & $\bm{\HeHp + \gamma \to \Hep + \mH}$ & $273518T_{\rm R}^{0.623525}\exp\left(-\frac{144044.0}{T_{\rm R}}\right)$ &  & 31 \\

\hypertarget{He_15}{He\_15} &  $\bm{\He + \me  \rightarrow  \Hep + 2\me}$ &
$\exp[-4.409864886 \times 10^{1}$ & & 14 \\
& & $\phantom{\exp[} \mbox{} + 2.391596563 \times 10^{1} \ln T_{\rm eV}$ & & \\
& & $\phantom{\exp[} \mbox{} - 1.07532302 \times 10^{1} (\ln T_{\rm eV})^{2}$ & & \\
& & $\phantom{\exp[} \mbox{} +3.05803875 \times 10^{0} (\ln T_{\rm eV})^{3}$ & & \\
& & $\phantom{\exp[} \mbox{} - 5.68511890 \times 10^{-1} (\ln T_{\rm eV})^{4}$ & & \\
& & $\phantom{\exp[} \mbox{} +6.79539123 \times 10^{-2} (\ln T_{\rm eV})^{5}$ & & \\
& & $\phantom{\exp[} \mbox{} -5.00905610 \times 10^{-3} (\ln T_{\rm eV})^{6}$ & & \\
& & $\phantom{\exp[} \mbox{} + 2.06723616\times 10^{-4}  (\ln T_{\rm eV})^{7}$ & & \\
& & $\phantom{\exp[} \mbox{} - 3.64916141 \times 10^{-6} (\ln T_{\rm eV})^{8}]$ & & \\
\hypertarget{He_16}{He\_16} & $\bm{\Hep + \me \rightarrow \Hepp + 2\me}$ &
$\exp[-6.87104099 \times 10^{1}$ & & 14 \\
& & $\phantom{\exp[} \mbox{} + 4.393347633 \times 10^{1} \ln T_{\rm eV}$ & & \\
& & $\phantom{\exp[} \mbox{} - 1.84806699 \times 10^{1} (\ln T_{\rm eV})^{2}$ & & \\
& & $\phantom{\exp[} \mbox{} + 4.70162649 \times 10^{0} (\ln T_{\rm eV})^{3}$ & & \\
& & $\phantom{\exp[} \mbox{} - 7.6924663 \times 10^{-1} (\ln T_{\rm eV})^{4}$ & & \\
& & $\phantom{\exp[} \mbox{} + 8.113042 \times 10^{-2} (\ln T_{\rm eV})^{5}$ & & \\
& & $\phantom{\exp[} \mbox{} - 5.32402063 \times 10^{-3} (\ln T_{\rm eV})^{6}$ & & \\
& & $\phantom{\exp[} \mbox{} + 1.97570531\times 10^{-4} (\ln T_{\rm eV})^{7}$ & & \\
& & $\phantom{\exp[} \mbox{} - 3.16558106\times 10^{-6} (\ln T_{\rm eV})^{8}$ & & \\

\hypertarget{He_17}{He\_17} & $\bm{\Hep + \Hm \rightarrow \He + \mH}$ &
$2.32 \times 10^{-7} \left(\frac{T}{300}\right)^{-0.52} \expf{}{T}{22400}$  & & 36 \\

\hypertarget{He_18}{He\_18} & $\bm{\He + \Hp + \gamma \to \HeHp + 2\gamma}$ & $0.130\times10^{-20} \left(\frac{T}{300}\right)^{0.408}\exp\left(-\frac{T}{36400}\right) $ &  & 35 \\
 &  & $ + 7.89\times10^{-20}\left(\frac{T}{300}\right)^{-0.399}\exp\left(-\frac{921}{T}\right)$ &  & \\

\hypertarget{He_19}{He\_19} & $\mHt + \Hep \rightarrow \He + \mH + \Hp$ &
$10^{-13}\cdot\left(-0.4732142857+0.0369047619\cdot T\right.$ & $T<78$~K & 37 \\
& & $ \phantom{10^{-13}\cdot(} \left.-0.1488095\cdot 10^{-3}\cdot T^2\right)$ & & \\
& & $10^{-13}\cdot\left(1.623809524-0.1587301587\cdot 10^{-2}\cdot T\right)$ & $78\,{\rm K}\le T<330$~K &  \\
& & $10^{-13}\cdot\left(0.111111524+0.1555555413\cdot 10^{-2}\cdot T\right.$ & $T\ge 330$~K & \\
& & $ \phantom{10^{-13}\cdot(} \left.+4.444444\cdot 10^{-6}\cdot T^2\right)$ & & \\

\hypertarget{He_20}{He\_20} & $\mHt + \Hep \rightarrow \mHtp + \He$ & $7.2 \times 10^{-15}$ & & 38 \\
\hypertarget{He_21}{He\_21} & $\He + \Hm \rightarrow \He + \mH + \me$ &
$4.1 \times 10^{-17} T^{2} \expf{-}{19870}{T} $ & & 39 \\

\hypertarget{He_22}{He\_22} & ${\Hep + \mHt \rightarrow \Hep + 2\mH}$ & $0.3\times 10^{-10}\left(\frac{T}{300}\right)^{0.5}\exp\left(-\frac{52000}{T}\right)$  &  & 26 \\

\hline

\hypertarget{D_01}{D\_01} & $\bm{\Dp + \me \rightarrow \mD  + \gamma}$ & the same as for \hyperlink{H_01}{\color{blue}H\_01} & &  \\
\hypertarget{D_02}{D\_02} & $\bm{\mD + \gamma \rightarrow \Dp + \me}$ & the same as for \hyperlink{H_02}{\color{blue}H\_02} & &  \\
\hypertarget{D_03}{D\_03}  &  $\bm{\mD + \Hp \rightarrow  \mH + \Dp}$ &
$2.0 \times 10^{-10} T^{0.402}  \exp \left(-\frac{37.1}{T} \right)$ &
$T\le 2 \cdot 10^{5} \: {\rm K}$ & 40 \\
& & $\phantom{=} \mbox{}- 3.31 \times 10^{-17} T^{1.48}$ & & \\
& & $\phantom{} 3.44 \times 10^{-10} T^{0.35}$ & $T > 2 \cdot 10^{5} \: {\rm K}$ & \\
\hypertarget{D_04}{D\_04} &  $\bm{\mH + \Dp  \rightarrow  \mD + \Hp}$ &
$2.06 \times 10^{-10} T^{0.396}  \exp \left(-\frac{33}{T} \right)$ & & 40 \\
& & $\phantom{ =}\mbox{} + 2.03 \times 10^{-9} T^{-0.332}$ & & \\

\hypertarget{D_05}{D\_05} & $\bm{\mHt + \Dp \rightarrow \hd + \Hp}$ &
$\left[0.417 + 0.846 \log{T} - 0.137 (\log{T})^{2} \right] \times 10^{-9}$ & & 41 \\

\hypertarget{D_06}{D\_06} & $\bm{\hd + \Hp \rightarrow \mHt + \Dp}$ & $1.1 \times 10^{-9} \expf{-}{488}{T}$ & & 41 \\
\hypertarget{D_07}{D\_07} & $\bm{\mH + \mD \rightarrow \hd + \gamma}$ &
$10^{-25} [2.80202 - 6.63697 \ln T $ & $T \le 200 \: {\rm K}$ & 42 \\
& & $\phantom{10^{-25} [} \mbox{} + 4.75619 (\ln T)^{2} -1.39325 (\ln T)^{3} $ & & \\
& & $\phantom{10^{-25} [}  \mbox{}  + 0.178259 (\ln T)^{4} - 0.00817097 (\ln T)^{5} ]$ & & \\
& & $\phantom{} 10^{-25} \exp [507.207 - 370.889 \ln T $ & $T > 200 \: {\rm K}$ & \\
& & $\phantom{10^{-25} \exp }  \mbox{} + 104.854 (\ln T)^2 - 14.4192 (\ln T)^{3} $ & & \\
& & $\phantom{10^{-25} \exp } \mbox{} + 0.971469 (\ln T)^{4} - 0.0258076 (\ln T)^{5} ]$ & & \\

\hypertarget{D_08}{D\_08} & $\bm{\mHt + \mD \rightarrow \hd + \mH}$ &
${\rm dex}\left[-56.4737 + 5.88886\log{T} \right.$ & $T \le 2000 \: {\rm K}$ & 43 \\
& & $\phantom{{\rm dex}[} \mbox{} + 7.19692 (\log{T})^{2}$ & & \\
& & $\phantom{{\rm dex}[} \mbox{} + 2.25069 (\log{T})^{3}$ & & \\
& & $\phantom{{\rm dex}[} \mbox{} - 2.16903  (\log{T})^{4}$ & & \\
& & $\left. \phantom{{\rm dex}[} \mbox{} + 0.317887 (\log{T})^{5} \right]$ & & \\
& & $\phantom{} 3.17 \times 10^{-10} \expf{-}{5207}{T}$ & $T > 2000 \: {\rm K}$ & \\
\hypertarget{D_09}{D\_09} & $\bm{\hdp + \mH \rightarrow \hd + \Hp}$ & the same as for \hyperlink{H_09}{\color{blue}H\_09} & &  \\
\hypertarget{D_10}{D\_10} &  $\bm{\hd + \mH  \rightarrow  \mHt + \mD}$ &
$5.25 \times 10^{-11} \expf{-}{4430}{T}$ & $T \le 200 \: {\rm K}$ & 44 \\
& & $\phantom{=} 5.25 \times 10^{-11} \exp \left(-\frac{4430}{T} +
 \frac{173900}{T^{2}}\right)$ & $T > 200 \: {\rm K}$ & \\

\hypertarget{D_11}{D\_11} & $\bm{\mD + \Hp \rightarrow \hdp + \gamma}$ & $3.9 \times 10^{-19}
\left(\frac{T}{300}\right)^{1.8} \expf{}{20}{T}$ & & 45 \\
\hypertarget{D_12}{D\_12} & $\bm{\mH + \Dp \rightarrow \hdp + \gamma}$ & $3.9 \times 10^{-19}
\left(\frac{T}{300}\right)^{1.8} \expf{}{20}{T}$ & & 45 \\

\hypertarget{D_13}{D\_13} & $\bm{\hdp + \gamma \to \mH + \Dp}$ & the same as for \hyperlink{H_08}{\color{blue}H\_08}  & & \\

\hypertarget{D_14}{D\_14} & $\bm{\hdp + \me \rightarrow \mH + \mD}$ & $7.2 \times 10^{-8} T^{-0.5}$ & & 46 \\

\hypertarget{D_15}{D\_15} & $\bm{\hdp + \mHt \to \htdp + \mH}$ & the same as for \hyperlink{H_13}{\color{blue}H\_13} & &  \\

\hypertarget{D_16}{D\_16} & $\bm{\mD + \htp \to \htdp + \mH}$ & $4.55\times 10^{-10} \left(\frac{T}{300}\right)^{-0.5} \expf{-}{900}{T}$ & & 28 \\

\hypertarget{D_17}{D\_17} & $\bm{\htdp + \me \to 2\mH + \mD}$ & $0.73\times 10^{-6}/\sqrt{T}$ & & 47 \\
\hypertarget{D_18}{D\_18} & $\bm{\htdp + \me \to \mHt + \mD}$ & $0.07\times 10^{-6}/\sqrt{T}$ & & 47 \\
\hypertarget{D_19}{D\_19} & $\bm{\htdp + \me \to \hd + \mH}$ & $0.20\times 10^{-6}/\sqrt{T}$ & & 47 \\

\hypertarget{D_20}{D\_20} & $\bm{\htdp + \mH \to \htp + \mD}$ & $2.0\times 10^{-8}T^{-1}\exp\left(-\frac{632}{T}\right)$ & &  58\\

\hypertarget{D_21}{D\_21} & $\bm{\hdp + \mH \rightarrow \mHtp + \mD}$  & $1.0 \times 10^{-9} \expf{-}{154}{T}$ & & 48 \\
\hypertarget{D_22}{D\_22}  &  $\bm{\mD + \Hm  \rightarrow  \hd + \me}$ & $7.5\times 10^{-10}$ & $T<300$ & 49 \\
&  & $2.0\times 10^{-9}T^{-0.17}$ & $T\ge 300$ & 49 \\
\hypertarget{D_23}{D\_23}  &  $\bm{\mH + \Dm  \rightarrow  \hd + \me}$ & $7.5\times 10^{-10}$ & $T<300$ & 49 \\
&  & $2.0\times 10^{-9}T^{-0.17}$ & $T\ge 300$ & 49 \\

\hypertarget{D_24}{D\_24} & $\bm{\hd + \Hp  \rightarrow \mHtp + \mD}$ &
$1.0 \times 10^{-9} \expf{-}{21600}{T}$ & & 50 \\

\hypertarget{D_25}{D\_25} & $\bm{\hdp + \mH \rightarrow \mHt + \Dp}$ & $1.0 \times 10^{-9}$ & & 50 \\
\hypertarget{D_26}{D\_26} & $\bm{\hdp + \gamma \to \Hp + \mD }$ & the same as for \hyperlink{H_08}{\color{blue}H\_08}  & & \\

\hypertarget{D_27}{D\_27} & $\bm{\mHtp + \mD \rightarrow \hdp + \mH}$ &
$1.07 \times 10^{-9} \left(\frac{T}{300}\right)^{0.062} \expf{-}{T}{41400}$ & & 51 \\

\hypertarget{D_28}{D\_28}  &  $\bm{\mD + \me  \rightarrow  \Dm + \gamma}$ & ${\rm dex}[-17.845 + 0.762 \log{T}$ & $T \le 6000 \: {\rm K}$ & 2 \\
& & $\phantom{ {\rm dex}[} \mbox{}+ 0.1523 (\log{T})^{2}$ & & \\
& & $\phantom{ {\rm dex}[} \mbox{}- 0.03274 (\log{T})^{3}] $ & & \\
& & $ {\rm dex}[-16.4199 + 0.1998 (\log{T})^{2}$ & $T > 6000 \: {\rm K}$ & \\
& & $ \phantom{ {\rm dex}[} \mbox{}-5.447  \times 10^{-3}  (\log{T})^{4}$ & & \\
& & $ \phantom{ {\rm dex}[} \mbox{}+ 4.0415 \times 10^{-5} (\log{T})^{6}]$  & & \\
\hypertarget{D_29}{D\_29}  &  $\bm{\mH + \Dm  \rightarrow  \mD + \Hm}$ &
$6.4 \times 10^{-9} \left(\frac{T}{300}\right)^{0.41}$ & & 48 \\
\hypertarget{D_30}{D\_30} &  $\bm{\mD + \Hm  \rightarrow  \mH + \Dm}$ &
$6.4 \times 10^{-9} \left(\frac{T}{300}\right)^{0.41}$ & & 48 \\
\hypertarget{D_31}{D\_31} & $\bm{\mD + \Dm \rightarrow \DD + \me}$ & the same as for \hyperlink{H_05}{\color{blue}H\_05} &  &  \\

\hypertarget{D_32}{D\_32} & $\bm{\Hp + \Dm \rightarrow \hdp + \me}$ &
$1.1 \times 10^{-9} \left(\frac{T}{300}\right)^{-0.4}$ & & 45 \\
\hypertarget{D_33}{D\_33} & $\bm{\Dp + \Hm \rightarrow \hdp + \me}$ &
$1.1 \times 10^{-9} \left(\frac{T}{300}\right)^{-0.4}$ & & 45 \\
\hypertarget{D_34}{D\_34} & $\bm{\Dp + \Dm \rightarrow \ddp + \me}$ &
$1.3 \times 10^{-9} \left(\frac{T}{300}\right)^{-0.4}$ & & 45 \\

\hypertarget{D_35}{D\_35}  &  $\bm{\Hp + \Dm  \rightarrow  \mD + \mH}$ & $2.4\times 10^{-6}(1.0+T/20000)/\sqrt{T}$  & & 52 \\

\hypertarget{D_36}{D\_36} & $\bm{\mD + \Dp \rightarrow \ddp + \gamma}$ &
$1.9 \times 10^{-19} T_{3}^{1.8} \expf{}{20}{T}$ & & 45 \\
\hypertarget{D_37}{D\_37}  &  $\bm{\mD + \mHtp  \rightarrow  \mHt + \Dp}$ & the same as for \hyperlink{H_09}{\color{blue}H\_09} & &  \\
\hypertarget{D_38}{D\_38} & $\bm{\mHtp + \mD \rightarrow \hd + \Hp}$ & $1.0 \times 10^{-9}$ & & 50 \\
\hypertarget{D_39}{D\_39} & $\bm{\hdp + \mD \rightarrow \ddp + \mH}$ & $1.0 \times 10^{-9}$ & & 53 \\
\hypertarget{D_40}{D\_40} & $\bm{\hdp + \mD \rightarrow \DD + \Hp}$ & $1.0 \times 10^{-9}$ & & 50 \\

\hypertarget{D_41}{D\_41}  &  $\bm{\mH + \ddp  \rightarrow  \DD + \Hp}$ & the same as for \hyperlink{H_09}{\color{blue}H\_09} & &  \\
\hypertarget{D_42}{D\_42} & $\bm{\ddp + \mH \rightarrow \hdp + \mD}$ & $1.0 \times 10^{-9} \expf{-}{472}{T}$ & & 53 \\
\hypertarget{D_43}{D\_43} & $\bm{\ddp + \mH \rightarrow \hd + \Dp}$ & $1.0 \times 10^{-9}$  & & 50 \\

\hypertarget{D_44}{D\_44} & $\bm{\hd + \Dp \rightarrow \DD + \Hp}$ & $1.0 \times 10^{-9}$ & & 53 \\

\hypertarget{D_45}{D\_45} & $\bm{\DD + \Hp \rightarrow \hdp + \mD}$ &
$ \left[ 5.18 \times 10^{-11} + 3.05 \times 10^{-9} \left(\frac{T}{10000}\right) \right.$ & & 54 \\
& & $\phantom{[} \left. \mbox{} -5.42 \times 10^{-10} \left(\frac{T}{10000}\right)^{2}
\right]\expf{-}{20100}{T}$ & & \\
\hypertarget{D_46}{D\_46} & $\bm{\DD + \Hp \rightarrow \ddp + \mH}$ & the same as for \hyperlink{H_10}{\color{blue}H\_10} &  & 52 \\

\hypertarget{D_47}{D\_47} & $\bm{\hd + \mD \rightarrow \DD + \mH}$ &
$1.15 \times 10^{-11} \expf{-}{3220}{T}$ & & 44 \\
\hypertarget{D_48}{D\_48} & $\bm{\DD + \mH \rightarrow \hd + \mD}$ &
${\rm dex}\left[-86.1558 + 4.53978 \log{T} \right.$ & $T \le 2200 \: {\rm K}$ &  43 \\
& & $\phantom{{\rm dex}[} \mbox{} + 33.5707 (\log{T})^{2}$ & & \\
& & $\phantom{{\rm dex}[} \mbox{} - 13.0449 (\log{T})^{3}$ & & \\
& & $\phantom{{\rm dex}[} \mbox{} + 1.22017  (\log{T})^{4}$ & & \\
& & $\left. \phantom{{\rm dex}[} \mbox{} + 0.0482453 (\log{T})^{5} \right]$ & & \\
& & $\phantom{} 2.67 \times 10^{-10} \expf{-}{5945}{T}$ & $T > 2200 \: {\rm K}$ & \\

\hypertarget{D_49}{D\_49} & $\bm{\hd + \gamma \to \mH + \mD}$ & $1.19\times 10^7 T_{\rm R}^{0.28}\exp\left(-\frac{145310}{T_{\rm R}}\right)$ & & 55\\

\hypertarget{D_50}{D\_50} & $\bm{\ddp + \gamma \rightarrow \mD + \Dp}$ &
the same as for \hyperlink{H_08}{\color{blue}H\_08} & &  \\
\hypertarget{D_51}{D\_51} & $\bm{\DD + \gamma \rightarrow \ddp + \me}$ &
the same as for \hyperlink{H_14}{\color{blue}H\_14} & &  \\
\hypertarget{D_52}{D\_52} & $\bm{\DD + \gamma \rightarrow \mD + \mD}$ &
the same as for \hyperlink{H_15}{\color{blue}H\_15} & &  \\
\hypertarget{D_53}{D\_53} & $\bm{\Dm + \gamma \rightarrow \mD + \me}$ & the same as for \hyperlink{H_04}{\color{blue}H\_04} & &  \\

\hypertarget{D_54}{D\_54} & $\bm{\hd + \htp \to \mHt + \htdp}$ & $1.0\times 10^{-9}\left(2.1-0.4\cdot \log(T)\right)$ & & 3\\

\hypertarget{D_55}{D\_55} & $\DD + \me \rightarrow \mD + \mD + \me$ &
$8.24 \times 10^{-9} T^{0.126} \expf{-}{105388}{T}$ & $v=0$ & 20 \\
& & $\phantom{} 2.75 \times 10^{-9} T^{0.163} \expf{-}{53339.7}{T}$ & LTE & \\
\hypertarget{D_56}{D\_56} & $\hdp + \mHt \to  \htp + \mD$ & the same as for \hyperlink{H_13}{\color{blue}H\_13} & & \\
\hypertarget{D_57}{D\_57} & $\htdp + \mHt \to \htp + \hd$ & $4.7\times 10^{-9}\exp\left(-\frac{215}{T}\right)$ & $T<100$~K & 3 \\
& & $5.5\times 10^{-10}$ & $T\ge 100$~K & \\
\hypertarget{D_58}{D\_58} & $\mD + \me \rightarrow \Dp + \me + \me$ & the same as for \hyperlink{H_20}{\color{blue}H\_20} & & 52 \\
\hypertarget{D_59}{D\_59} &  $\hd + \me  \rightarrow  \mD + \Hm $ &
$1.35 \times 10^{-9} T^{-1.27} \expf{-}{43000}{T}$ & & 56 \\
\hypertarget{D_60}{D\_60} &  $\hd + \me  \rightarrow  \mH + \Dm $ &
$1.35 \times 10^{-9} T^{-1.27} \expf{-}{43000}{T}$ & & 56 \\
\hypertarget{D_61}{D\_61} & $\hd + \me \rightarrow \mH + \mD + \me$ &
$5.09 \times 10^{-9} T^{0.128}  \expf{-}{103258}{T}$ & $v=0$ & 57 \\
& & $\phantom{} 1.04 \times 10^{-9} T^{0.218}  \expf{-}{53070.7}{T}$ & LTE & \\
\hypertarget{D_62}{D\_62}  &  $\mHt + \Dp  \rightarrow  \mHtp + \mD $ & the same as for \hyperlink{H_10}{\color{blue}H\_10} & & 52 \\
\hypertarget{D_63}{D\_63} & $\mHt + \Dp \rightarrow \hdp + \mH$ &
$\left[ 1.04 \times 10^{-9} + 9.52 \times 10^{-9} \left(\frac{T}{10000}\right) \right.$ & & 54 \\
& & $\phantom{=} \left. \mbox{} -1.81 \times 10^{-9} \left(\frac{T}{10000}\right)^{2}
\right]\expf{-}{21000}{T}$ & & \\
\hypertarget{D_64}{D\_64} & $\hd + \Hp \rightarrow \hdp + \mH$ & the same as for \hyperlink{H_10}{\color{blue}H\_10} & & 52 \\

\hypertarget{D_65}{D\_65} & $\hd + \mH \rightarrow \mH + \mD + \mH$ & the same as for \hyperlink{H_19}{\color{blue}H\_19} & & 18 \\

\hypertarget{D_66}{D\_66} & $\hd + \He \rightarrow \mH + \mD + \He$ & the same as for \hyperlink{H_30}{\color{blue}H\_30} & & 18 \\

\hypertarget{D_67}{D\_67} & $\hd + \Hep \rightarrow \hdp + \He$ & the same as for \hyperlink{He_20}{\color{blue}He\_20} & & 52 \\
\hypertarget{D_68}{D\_68}  &  $\hd + \Hep  \rightarrow  \He + \Hp + \mD $ &
$1.85 \times 10^{-14} \expf{}{35}{T}$ & & 49 \\
\hypertarget{D_69}{D\_69}  &  $\hd + \Hep  \rightarrow  \He + \mH + \Dp $ &
$1.85 \times 10^{-14} \expf{}{35}{T}$ & & 49 \\

\hypertarget{D_70}{D\_70} & $\hdp + \mD \rightarrow \hd + \Dp$  & the same as for \hyperlink{H_09}{\color{blue}H\_09} & & 52 \\

\hypertarget{D_71}{D\_71} &  $\DD + \me  \rightarrow  \mD + \Dm $ &
$6.7 \times 10^{-11} T^{-1.27} \expf{-}{43000}{T}$ & & 56 \\

\hypertarget{D_72}{D\_72}  &  $\Dm + \me  \rightarrow  \mD + \me + \me $ & the same as for \hyperlink{H_31}{\color{blue}H\_31} & & 52 \\
\hypertarget{D_73}{D\_73}  &  $\Dm + \mH  \rightarrow  \mD + \mH + \me $ &  the same as for \hyperlink{H_32}{\color{blue}H\_32} & & 52 \\

\hypertarget{D_74}{D\_74} &  $\Dp + \Hm  \rightarrow  \mD + \mH $ & the same as for \hyperlink{H_06}{\color{blue}H\_06} & & 52 \\

\hypertarget{D_75}{D\_75}  &  $\Dp + \Dm  \rightarrow  \mD + \mD $ & the same as for \hyperlink{H_06}{\color{blue}H\_06}  & & 52 \\
\hypertarget{D_76}{D\_76} & $\mHtp + \Dm \rightarrow \mHt  + \mD$ &
$1.7 \times 10^{-7} \left(\frac{T}{300}\right)^{-0.5}$ & & 45 \\
\hypertarget{D_77}{D\_77} & $\mHtp + \Dm \rightarrow \mH + \mH  + \mD$ &
$1.7 \times 10^{-7} \left(\frac{T}{300}\right)^{-0.5}$ & & 45 \\
\hypertarget{D_78}{D\_78} & $\hdp + \Hm \rightarrow \hd  + \mH$ &
$1.5 \times 10^{-7} \left(\frac{T}{300}\right)^{-0.5}$ & & 45 \\
\hypertarget{D_79}{D\_79} & $\hdp + \Hm \rightarrow \mD + \mH + \mH$ &
$1.5 \times 10^{-7} \left(\frac{T}{300}\right)^{-0.5}$ & & 45 \\
\hypertarget{D_80}{D\_80} & $\hdp + \Dm \rightarrow \hd  + \mD$ &
$1.9 \times 10^{-7} \left(\frac{T}{300}\right)^{-0.5}$ & & 45 \\
\hypertarget{D_81}{D\_81} & $\hdp + \Dm \rightarrow \mD + \mH + \mD$ &
$1.9 \times 10^{-7} \left(\frac{T}{300}\right)^{-0.5}$ & & 45 \\
\hypertarget{D_82}{D\_82} & $\ddp + \Hm \rightarrow \DD + \mH$ &
$1.5 \times 10^{-7} \left(\frac{T}{300}\right)^{-0.5}$ & & 45 \\
\hypertarget{D_83}{D\_83} & $\ddp + \Hm \rightarrow \mD + \mD + \mH$ &
$1.5 \times 10^{-7} \left(\frac{T}{300}\right)^{-0.5}$  & & 45 \\
\hypertarget{D_84}{D\_84} & $\ddp + \Dm \rightarrow \DD + \mD$ &
$2.0 \times 10^{-7} \left(\frac{T}{300}\right)^{-0.5}$ & & 45 \\
\hypertarget{D_85}{D\_85} & $\ddp + \Dm \rightarrow \mD + \mD + \mD$ &
$2.0 \times 10^{-7} \left(\frac{T}{300}\right)^{-0.5}$ & & 45 \\

\hypertarget{D_86}{D\_86}  &  $\mD + \ddp  \rightarrow  \DD + \Dp $ & the same as for \hyperlink{H_09}{\color{blue}H\_09} & & 52 \\

\hypertarget{D_87}{D\_87} & $\hd + \Dp \rightarrow \hdp + \mD$ & the same as for \hyperlink{H_10}{\color{blue}H\_10} & & 52 \\

\hypertarget{D_88}{D\_88} & $\hd + \Dp \rightarrow \ddp + \mH$ &
$\left[ 3.54 \times 10^{-9} + 7.50 \times 10^{-10} \left(\frac{T}{10000}\right) \right.$ & & 54 \\
& & $\phantom{[} \left. \mbox{} -2.92 \times 10^{-10} \left(\frac{T}{10000}\right)^{2}
\right]\expf{-}{21100}{T}$ & & \\
\hypertarget{D_89}{D\_89} & $\DD + \Hp \rightarrow \hd + \Dp$ & $2.1 \times 10^{-9} \expf{-}{491}{T}$ & & 53 \\
\hypertarget{D_90}{D\_90} & $\DD + \Dp \rightarrow \ddp + \mD$ & the same as for \hyperlink{H_10}{\color{blue}H\_10} & & 52 \\
\hypertarget{D_91}{D\_91} & $\hd + \mHt \rightarrow \mH + \mD + \mHt$ & the same as for \hyperlink{H_29}{\color{blue}H\_29} but with $n_{cr}\to 100n_{cr}$ & & 18 \\
\hypertarget{D_92}{D\_92} & $\DD + \mH \rightarrow \mD + \mD + \mH$ & the same as for \hyperlink{H_19}{\color{blue}H\_19} & & 52 \\
\hypertarget{D_93}{D\_93} & $\DD + \mHt \rightarrow \mD + \mD + \mHt$ & the same as for \hyperlink{H_29}{\color{blue}H\_29} & & 52 \\
\hypertarget{D_94}{D\_94} & $\He + \Dp \rightarrow \mD + \Hep$ &
$1.85 \times 10^{-9} T^{-0.75} \expf{-}{127500}{T}$ &
$ T \leq 10000 \: {\rm K}$ & 45 \\
& & $\phantom{} 5.9 \times 10^{-37} T^{4.74}$ & $T > 10000 \: {\rm K}$ & \\
\hypertarget{D_95}{D\_95} & $\Hep + \mD \rightarrow \Dp + \He$ &
$1.1 \times 10^{-15} \left(\frac{T}{300}\right)^{0.25}$ & & 45 \\
\hypertarget{D_96}{D\_96} & $\Dm + \He  \rightarrow  \mD + \He + \me $ &
$1.5 \times 10^{-17} T^{2} \expf{-}{19870}{T}$ & & 45 \\
\hypertarget{D_97}{D\_97} &  $\Hep + \Dm  \rightarrow  \He + \mD $ &
$3.03 \times 10^{-7} \left(\frac{T}{300}\right)^{-0.52} \expf{}{T}{22400}$ & & 45 \\
\hypertarget{D_98}{D\_98} & $\DD + \Hep \rightarrow \ddp + \He$ & $2.5 \times 10^{-14}$ & & 53 \\
\hypertarget{D_99}{D\_99}  &  $\DD + \Hep  \rightarrow  \He + \Dp + \mD $ &
$1.1 \times 10^{-13} T_{3}^{-0.24}$ & & 53 \\
\hypertarget{D_100}{D\_100} & $\DD + \He \rightarrow \mD + \mD + \He$ & the same as for \hyperlink{H_30}{\color{blue}H\_30} & & 52 \\
\end{longtable}
\medskip

{\bf Note}: $T$ and $T_{\rm eV}$ are the gas temperature in units of K and eV respectively. References are to the primary source of data for each
reaction. Reactions with a peak contribution to the synthesis or destruction of any reactant exceeding 0.1 \% are bolded.\\
{\bf References}:

1: \citet{Seager1999},
2: \citet{WIS79},
3: \citet{Galli1998},
4: \citet{cdg99},
5: \citet{RAM76},
6: \citet{KAR79},
7: \citet{SAV04},
8: \citet{POU78},
9: \citet{Coppola2011a},
10: \citet{Theard1974},
11: \citet{Novosyadlyj2022},
 12: \citet{MAC86},
 13:  \citet{ls83},
 14: \citet{JAN87},
 15: this paper: Fit based on cross-section from https://home.strw.leidenuniv.nl/~ewine/photo/,
16: \citet{Mendez2006},
17: \citet{Janev2003},
18: \citet{Glover2008},
19: \citet{sa67},
20: \citet{tt02a},
 21: \citet{MAR98},
22: \citet{sk87},
23: \citet{drcm87},
24: \citet{dl87},
25: \citet{wk75},
26: \citet{Albertsson2014},
27: \citet{Soon1992},
28:  Database of kinetic data of interest for astrochemical (KIDA, kida.astrochem-tools.org),
29: \citet{Courtney2021},
30: \citet{Fazio:2014},
31: \citet{Coppola2017},
32: \citet{FER92},
33: \citet{kldd93},
34: \citet{z89},
35: \citet{Black1978},
36: \citet{ph94},
37: this paper: Fit to data from \citet{Schauer1989, Johnsen1980}
38: \citet{Glover2008}: Barlow S. G. , 1984, PhD thesis, Univ. Colorado
39: \citet{h82},
40: \citet{sav02},
41: \citet{ger82},
42: \citet{dic05},
43: Fit of \citet{Glover2008} to data from \citet{mie03},
44: \citet{s59},
45: Same as corresponding H reaction, but scaled by D reduced mass (see \citet{Glover2008}),
46: \citet{sss95},
47: \citet{Datz1995,Larsson1996},
48: \citet{dm56}, scaled by D reduced mass (see \citet{Glover2008}),
49: Same as corresponding H reaction, with branching ratio assumed uniform (see \citet{Glover2008}),
50: estimate from \citet{Glover2008},
51: \citet{ljb95},
52: Same as corresponding H reaction (see \citet{Glover2008}),
53: \citet{wfp04},
54: Fit of \citet{Glover2008} based on cross-section from \citet{ws02}
55: \citet{Coppola2011b},
56: \citet{xf01},
57: \citet{tt02b},
58: \citet{Adams1985},
59: \citet{Stancil1998}.

\tiny
\begin{table}[]
\caption{Map of chemical reactions.}
\vspace{-0.5cm}
 \label{tab:1}
\begin{center}
\resizebox{0.74 \textheight}{!}{
\renewcommand{\arraystretch}{1.1}
\setlength{\arrayrulewidth}{1.0pt}
\begin{tabular}{| >{\arraybackslash\hspace{0pt}}p{0.04\textwidth}||
             *{20}{>{\centering\arraybackslash\hspace{0pt}}p{0.05\textwidth}  |} }
\hline
     & $\rm e$ & $\rm p$ & $\rm H$ & $\rm H^{-}$ & $\rm H_2$ & $\rm H_2^{+}$ & $\rm H_3^{+}$ & $\rm He$ & $\rm He^{+}$ & $\rm He^{++}$ & $\rm HeH^{+}$ & $\rm D$ & $\rm D^{-}$ & $\rm D^{+}$ & $\rm D_2$ & $\rm D_2^{+}$ & $\rm HD$ & $\rm HD^{+}$ & $\rm H_2D^{+}$ & $\rm \gamma$\\
\hline
$\rm e$ & {--} & \hyperlink{H_02}{\color{blue}$\rm H\_02$} \hyperlink{H_20}{\color{cyan}$\rm H\_20$} \hyperlink{H_23}{\color{cyan}$\rm H\_23$} \hyperlink{H_26}{\color{cyan}$\rm H\_26$} \hyperlink{H_41}{\color{cyan}$\rm H\_41$} \hyperlink{H_42}{\color{cyan}$\rm H\_42$}  & \hyperlink{H_04}{\color{blue}$\rm H\_04$} \hyperlink{H_28}{\color{cyan}$\rm H\_28$} \hyperlink{H_31}{\color{cyan}$\rm H\_31$} \hyperlink{H_32}{\color{cyan}$\rm H\_32$} \hyperlink{H_40}{\color{cyan}$\rm H\_40$} \hyperlink{H_41}{\color{cyan}$\rm H\_41$} \hyperlink{H_42}{\color{cyan}$\rm H\_42$} \hyperlink{He_21}{\color{cyan}$\rm He\_21$} \hyperlink{D_61}{\color{cyan}$\rm D\_61$} \hyperlink{D_73}{\color{cyan}$\rm D\_73$}  & {--} & \hyperlink{H_05}{\color{blue}$\rm H\_05$}  & \hyperlink{H_11}{\color{blue}$\rm H\_11$} \hyperlink{H_14}{\color{blue}$\rm H\_14$} \hyperlink{H_40}{\color{cyan}$\rm H\_40$}  & {--} & \hyperlink{He_21}{\color{cyan}$\rm He\_21$} \hyperlink{D_96}{\color{cyan}$\rm D\_96$}  & \hyperlink{He_02}{\color{blue}$\rm He\_02$} \hyperlink{He_15}{\color{cyan}$\rm He\_15$}  & \hyperlink{He_07}{\color{blue}$\rm He\_07$} \hyperlink{He_16}{\color{cyan}$\rm He\_16$}  & {--} & \hyperlink{D_53}{\color{blue}$\rm D\_53$} \hyperlink{D_54}{\color{cyan}$\rm D\_54$} \hyperlink{D_61}{\color{cyan}$\rm D\_61$} \hyperlink{D_72}{\color{cyan}$\rm D\_72$} \hyperlink{D_73}{\color{cyan}$\rm D\_73$} \hyperlink{D_96}{\color{cyan}$\rm D\_96$}  & {--} & \hyperlink{D_02}{\color{blue}$\rm D\_02$} \hyperlink{D_58}{\color{cyan}$\rm D\_58$}  & \hyperlink{D_31}{\color{blue}$\rm D\_31$}  & \hyperlink{D_34}{\color{blue}$\rm D\_34$} \hyperlink{D_51}{\color{blue}$\rm D\_51$}  & \hyperlink{D_22}{\color{blue}$\rm D\_22$} \hyperlink{D_23}{\color{blue}$\rm D\_23$}  & \hyperlink{D_32}{\color{blue}$\rm D\_32$} \hyperlink{D_33}{\color{blue}$\rm D\_33$}  & {--} & {--}\\
\hline
$\rm p$ & \hyperlink{H_01}{\color{red}$\rm H\_01$}  & \hyperlink{H_26}{\color{cyan}$\rm H\_26$}  & \hyperlink{H_08}{\color{blue}$\rm H\_08$} \hyperlink{H_37}{\color{cyan}$\rm H\_37$} \hyperlink{H_38}{\color{cyan}$\rm H\_38$} \hyperlink{H_41}{\color{cyan}$\rm H\_41$} \hyperlink{H_42}{\color{cyan}$\rm H\_42$} \hyperlink{He_19}{\color{cyan}$\rm He\_19$}  & {--} & \hyperlink{H_09}{\color{blue}$\rm H\_09$} \hyperlink{H_21}{\color{blue}$\rm H\_21$} \hyperlink{H_24}{\color{blue}$\rm H\_24$}  & {--} & {--} & \hyperlink{He_05}{\color{blue}$\rm He\_05$} \hyperlink{He_09}{\color{blue}$\rm He\_09$} \hyperlink{He_19}{\color{cyan}$\rm He\_19$} \hyperlink{D_68}{\color{cyan}$\rm D\_68$}  & {--} & {--} & {--} & \hyperlink{D_04}{\color{blue}$\rm D\_04$} \hyperlink{D_26}{\color{blue}$\rm D\_26$} \hyperlink{D_68}{\color{cyan}$\rm D\_68$}  & {--} & {--} & \hyperlink{D_40}{\color{blue}$\rm D\_40$} \hyperlink{D_41}{\color{blue}$\rm D\_41$} \hyperlink{D_44}{\color{blue}$\rm D\_44$}  & {--} & \hyperlink{D_05}{\color{blue}$\rm D\_05$} \hyperlink{D_09}{\color{blue}$\rm D\_09$} \hyperlink{D_38}{\color{blue}$\rm D\_38$}  & {--} & {--} & {--}\\
\hline
$\rm H$ & \hyperlink{H_03}{\color{red}$\rm H\_03$} \hyperlink{H_20}{\color{red}$\rm H\_20$} \hyperlink{H_23}{\color{red}$\rm H\_23$}  & \hyperlink{H_07}{\color{red}$\rm H\_07$} \hyperlink{H_24}{\color{orange}$\rm H\_24$} \hyperlink{H_43}{\color{orange}$\rm H\_43$}  & \hyperlink{H_06}{\color{blue}$\rm H\_06$} \hyperlink{H_12}{\color{blue}$\rm H\_12$} \hyperlink{H_15}{\color{blue}$\rm H\_15$} \hyperlink{H_19}{\color{cyan}$\rm H\_19$} \hyperlink{H_22}{\color{cyan}$\rm H\_22$} \hyperlink{H_25}{\color{orange}$\rm H\_25$} \hyperlink{H_28}{\color{cyan}$\rm H\_28$} \hyperlink{H_32}{\color{cyan}$\rm H\_32$} \hyperlink{H_34}{\color{cyan}$\rm H\_34$} \hyperlink{H_35}{\color{orange}$\rm H\_35$} \hyperlink{H_36}{\color{orange}$\rm H\_36$} \hyperlink{H_37}{\color{cyan}$\rm H\_37$} \hyperlink{H_38}{\color{cyan}$\rm H\_38$} \hyperlink{H_42}{\color{red}$\rm H\_42$} \hyperlink{D_17}{\color{cyan}$\rm D\_17$} \hyperlink{D_65}{\color{cyan}$\rm D\_65$} \hyperlink{D_77}{\color{cyan}$\rm D\_77$}  & \hyperlink{H_27}{\color{blue}$\rm H\_27$}  & \hyperlink{H_17}{\color{blue}$\rm H\_17$} \hyperlink{H_25}{\color{blue}$\rm H\_25$} \hyperlink{H_29}{\color{cyan}$\rm H\_29$} \hyperlink{H_33}{\color{blue}$\rm H\_33$} \hyperlink{D_91}{\color{cyan}$\rm D\_91$}  & \hyperlink{H_10}{\color{blue}$\rm H\_10$} \hyperlink{H_40}{\color{cyan}$\rm H\_40$} \hyperlink{H_43}{\color{blue}$\rm H\_43$}  & \hyperlink{H_13}{\color{blue}$\rm H\_13$} \hyperlink{H_39}{\color{cyan}$\rm H\_39$}  & \hyperlink{H_30}{\color{cyan}$\rm H\_30$} \hyperlink{He_12}{\color{blue}$\rm He\_12$} \hyperlink{He_17}{\color{blue}$\rm He\_17$} \hyperlink{He_19}{\color{cyan}$\rm He\_19$} \hyperlink{He_21}{\color{cyan}$\rm He\_21$} \hyperlink{D_66}{\color{cyan}$\rm D\_66$} \hyperlink{D_69}{\color{cyan}$\rm D\_69$}  & \hyperlink{He_08}{\color{blue}$\rm He\_08$} \hyperlink{He_14}{\color{blue}$\rm He\_14$} \hyperlink{He_22}{\color{cyan}$\rm He\_22$}  & {--} & \hyperlink{He_10}{\color{blue}$\rm He\_10$}  & \hyperlink{D_14}{\color{blue}$\rm D\_14$} \hyperlink{D_17}{\color{cyan}$\rm D\_17$} \hyperlink{D_35}{\color{blue}$\rm D\_35$} \hyperlink{D_49}{\color{blue}$\rm D\_49$} \hyperlink{D_61}{\color{cyan}$\rm D\_61$} \hyperlink{D_65}{\color{cyan}$\rm D\_65$} \hyperlink{D_66}{\color{cyan}$\rm D\_66$} \hyperlink{D_73}{\color{cyan}$\rm D\_73$} \hyperlink{D_74}{\color{blue}$\rm D\_74$} \hyperlink{D_77}{\color{cyan}$\rm D\_77$} \hyperlink{D_79}{\color{cyan}$\rm D\_79$} \hyperlink{D_81}{\color{cyan}$\rm D\_81$} \hyperlink{D_83}{\color{cyan}$\rm D\_83$} \hyperlink{D_91}{\color{cyan}$\rm D\_91$} \hyperlink{D_92}{\color{cyan}$\rm D\_92$}  & \hyperlink{D_30}{\color{blue}$\rm D\_30$} \hyperlink{D_60}{\color{blue}$\rm D\_60$}  & \hyperlink{D_03}{\color{blue}$\rm D\_03$} \hyperlink{D_13}{\color{blue}$\rm D\_13$} \hyperlink{D_69}{\color{cyan}$\rm D\_69$}  & \hyperlink{D_47}{\color{blue}$\rm D\_47$} \hyperlink{D_82}{\color{blue}$\rm D\_82$}  & \hyperlink{D_39}{\color{blue}$\rm D\_39$} \hyperlink{D_46}{\color{blue}$\rm D\_46$} \hyperlink{D_88}{\color{blue}$\rm D\_88$}  & \hyperlink{D_08}{\color{blue}$\rm D\_08$} \hyperlink{D_19}{\color{blue}$\rm D\_19$} \hyperlink{D_78}{\color{blue}$\rm D\_78$}  & \hyperlink{D_27}{\color{blue}$\rm D\_27$} \hyperlink{D_63}{\color{blue}$\rm D\_63$} \hyperlink{D_64}{\color{blue}$\rm D\_64$}  & \hyperlink{D_15}{\color{blue}$\rm D\_15$} \hyperlink{D_16}{\color{blue}$\rm D\_16$}  & \hyperlink{H_01}{\color{blue}$\rm H\_01$} \\
\hline
$\rm H^{-}$ & \hyperlink{H_31}{\color{red}$\rm H\_31$}  & \hyperlink{H_06}{\color{red}$\rm H\_06$} \hyperlink{H_11}{\color{red}$\rm H\_11$}  & \hyperlink{H_05}{\color{red}$\rm H\_05$} \hyperlink{H_32}{\color{red}$\rm H\_32$}  & {--} & {--} & {--} & {--} & {--} & {--} & {--} & {--} & \hyperlink{D_29}{\color{blue}$\rm D\_29$} \hyperlink{D_59}{\color{blue}$\rm D\_59$}  & {--} & {--} & {--} & {--} & {--} & {--} & {--} & \hyperlink{H_03}{\color{blue}$\rm H\_03$} \\
\hline
$\rm H_2$ & \hyperlink{H_27}{\color{red}$\rm H\_27$} \hyperlink{H_28}{\color{red}$\rm H\_28$}  & \hyperlink{H_10}{\color{red}$\rm H\_10$} \hyperlink{H_18}{\color{red}$\rm H\_18$} \hyperlink{H_38}{\color{red}$\rm H\_38$}  & \hyperlink{H_19}{\color{red}$\rm H\_19$} \hyperlink{H_35}{\color{orange}$\rm H\_35$}  & {--} & \hyperlink{H_29}{\color{red}$\rm H\_29$} \hyperlink{H_35}{\color{blue}$\rm H\_35$} \hyperlink{H_44}{\color{blue}$\rm H\_44$}  & \hyperlink{H_16}{\color{blue}$\rm H\_16$}  & {--} & \hyperlink{H_36}{\color{blue}$\rm H\_36$}  & {--} & {--} & {--} & \hyperlink{D_10}{\color{blue}$\rm D\_10$} \hyperlink{D_18}{\color{blue}$\rm D\_18$} \hyperlink{D_76}{\color{blue}$\rm D\_76$} \hyperlink{D_91}{\color{cyan}$\rm D\_91$} \hyperlink{D_93}{\color{cyan}$\rm D\_93$}  & {--} & \hyperlink{D_06}{\color{blue}$\rm D\_06$} \hyperlink{D_25}{\color{blue}$\rm D\_25$} \hyperlink{D_37}{\color{blue}$\rm D\_37$}  & {--} & {--} & {--} & {--} & \hyperlink{D_55}{\color{blue}$\rm D\_55$}  & {--}\\
\hline
$\rm H_2^{+}$ & \hyperlink{H_12}{\color{red}$\rm H\_12$} \hyperlink{H_41}{\color{red}$\rm H\_41$}  & {--} & \hyperlink{H_09}{\color{red}$\rm H\_09$} \hyperlink{H_37}{\color{red}$\rm H\_37$}  & \hyperlink{H_33}{\color{red}$\rm H\_33$} \hyperlink{H_34}{\color{red}$\rm H\_34$}  & \hyperlink{H_13}{\color{red}$\rm H\_13$}  & {--} & {--} & \hyperlink{He_04}{\color{blue}$\rm He\_04$} \hyperlink{He_20}{\color{blue}$\rm He\_20$}  & {--} & {--} & {--} & \hyperlink{D_21}{\color{blue}$\rm D\_21$} \hyperlink{D_24}{\color{blue}$\rm D\_24$} \hyperlink{D_62}{\color{blue}$\rm D\_62$}  & {--} & {--} & {--} & {--} & {--} & {--} & {--} & \hyperlink{H_07}{\color{blue}$\rm H\_07$} \\
\hline
$\rm H_3^{+}$ & \hyperlink{H_17}{\color{red}$\rm H\_17$} \hyperlink{H_22}{\color{red}$\rm H\_22$} \hyperlink{H_40}{\color{red}$\rm H\_40$}  & {--} & \hyperlink{H_16}{\color{red}$\rm H\_16$}  & \hyperlink{H_44}{\color{red}$\rm H\_44$}  & \hyperlink{H_39}{\color{red}$\rm H\_39$}  & {--} & {--} & \hyperlink{He_13}{\color{blue}$\rm He\_13$}  & {--} & {--} & {--} & \hyperlink{D_20}{\color{blue}$\rm D\_20$} \hyperlink{D_56}{\color{blue}$\rm D\_56$}  & {--} & {--} & {--} & {--} & \hyperlink{D_57}{\color{blue}$\rm D\_57$}  & {--} & {--} & \hyperlink{H_18}{\color{blue}$\rm H\_18$} \\
\hline
$\rm He$ & \hyperlink{He_15}{\color{red}$\rm He\_15$}  & \hyperlink{He_03}{\color{red}$\rm He\_03$} \hyperlink{He_08}{\color{red}$\rm He\_08$} \hyperlink{He_18}{\color{orange}$\rm He\_18$}  & \hyperlink{H_36}{\color{orange}$\rm H\_36$}  & \hyperlink{He_21}{\color{red}$\rm He\_21$}  & \hyperlink{H_30}{\color{red}$\rm H\_30$}  & \hyperlink{He_10}{\color{red}$\rm He\_10$}  & {--} & {--} & {--} & {--} & {--} & \hyperlink{D_66}{\color{cyan}$\rm D\_66$} \hyperlink{D_68}{\color{cyan}$\rm D\_68$} \hyperlink{D_96}{\color{cyan}$\rm D\_96$} \hyperlink{D_97}{\color{blue}$\rm D\_97$} \hyperlink{D_99}{\color{cyan}$\rm D\_99$} \hyperlink{D_100}{\color{cyan}$\rm D\_100$}  & {--} & \hyperlink{D_69}{\color{cyan}$\rm D\_69$} \hyperlink{D_95}{\color{blue}$\rm D\_95$} \hyperlink{D_99}{\color{cyan}$\rm D\_99$}  & {--} & \hyperlink{D_98}{\color{blue}$\rm D\_98$}  & {--} & \hyperlink{D_67}{\color{blue}$\rm D\_67$}  & {--} & \hyperlink{He_01}{\color{blue}$\rm He\_01$} \\
\hline
$\rm He^{+}$ & \hyperlink{He_01}{\color{red}$\rm He\_01$} \hyperlink{He_16}{\color{red}$\rm He\_16$}  & {--} & \hyperlink{He_09}{\color{red}$\rm He\_09$} \hyperlink{He_11}{\color{red}$\rm He\_11$}  & \hyperlink{He_17}{\color{red}$\rm He\_17$}  & \hyperlink{He_19}{\color{red}$\rm He\_19$} \hyperlink{He_20}{\color{red}$\rm He\_20$} \hyperlink{He_22}{\color{red}$\rm He\_22$}  & {--} & {--} & {--} & {--} & {--} & {--} & \hyperlink{D_94}{\color{blue}$\rm D\_94$}  & {--} & {--} & {--} & {--} & {--} & {--} & {--} & \hyperlink{He_06}{\color{blue}$\rm He\_06$} \\
\hline
$\rm He^{++}$ & \hyperlink{He_06}{\color{red}$\rm He\_06$}  & {--} & {--} & {--} & {--} & {--} & {--} & {--} & {--} & {--} & {--} & {--} & {--} & {--} & {--} & {--} & {--} & {--} & {--} & {--}\\
\hline
$\rm HeH^{+}$ & \hyperlink{He_12}{\color{red}$\rm He\_12$}  & {--} & \hyperlink{He_04}{\color{red}$\rm He\_04$}  & {--} & \hyperlink{He_13}{\color{red}$\rm He\_13$}  & {--} & {--} & {--} & {--} & {--} & {--} & {--} & {--} & {--} & {--} & {--} & {--} & {--} & {--} & \hyperlink{He_03}{\color{blue}$\rm He\_03$} \hyperlink{He_11}{\color{blue}$\rm He\_11$} \hyperlink{He_18}{\color{cyan}$\rm He\_18$} \\
\hline
$\rm D$ & \hyperlink{D_28}{\color{red}$\rm D\_28$} \hyperlink{D_58}{\color{red}$\rm D\_58$}  & \hyperlink{D_03}{\color{red}$\rm D\_03$} \hyperlink{D_11}{\color{red}$\rm D\_11$}  & \hyperlink{D_07}{\color{red}$\rm D\_07$}  & \hyperlink{D_22}{\color{red}$\rm D\_22$} \hyperlink{D_30}{\color{red}$\rm D\_30$}  & \hyperlink{D_08}{\color{red}$\rm D\_08$}  & \hyperlink{D_27}{\color{red}$\rm D\_27$} \hyperlink{D_37}{\color{red}$\rm D\_37$} \hyperlink{D_38}{\color{red}$\rm D\_38$}  & \hyperlink{D_16}{\color{red}$\rm D\_16$}  & {--} & \hyperlink{D_95}{\color{red}$\rm D\_95$}  & {--} & {--} & \hyperlink{D_52}{\color{blue}$\rm D\_52$} \hyperlink{D_54}{\color{cyan}$\rm D\_54$} \hyperlink{D_75}{\color{blue}$\rm D\_75$} \hyperlink{D_81}{\color{cyan}$\rm D\_81$} \hyperlink{D_83}{\color{cyan}$\rm D\_83$} \hyperlink{D_85}{\color{cyan}$\rm D\_85$} \hyperlink{D_92}{\color{cyan}$\rm D\_92$}  & \hyperlink{D_71}{\color{blue}$\rm D\_71$}  & \hyperlink{D_50}{\color{blue}$\rm D\_50$} \hyperlink{D_99}{\color{cyan}$\rm D\_99$}  & \hyperlink{D_84}{\color{blue}$\rm D\_84$}  & \hyperlink{D_90}{\color{blue}$\rm D\_90$}  & \hyperlink{D_48}{\color{blue}$\rm D\_48$} \hyperlink{D_80}{\color{blue}$\rm D\_80$}  & \hyperlink{D_42}{\color{blue}$\rm D\_42$} \hyperlink{D_45}{\color{blue}$\rm D\_45$} \hyperlink{D_87}{\color{blue}$\rm D\_87$}  & {--} & \hyperlink{D_01}{\color{blue}$\rm D\_01$} \\
\hline
$\rm D^{-}$ & \hyperlink{D_72}{\color{red}$\rm D\_72$}  & \hyperlink{D_32}{\color{red}$\rm D\_32$} \hyperlink{D_35}{\color{red}$\rm D\_35$}  & \hyperlink{D_23}{\color{red}$\rm D\_23$} \hyperlink{D_29}{\color{red}$\rm D\_29$} \hyperlink{D_73}{\color{red}$\rm D\_73$}  & {--} & {--} & \hyperlink{D_76}{\color{red}$\rm D\_76$} \hyperlink{D_77}{\color{red}$\rm D\_77$}  & {--} & \hyperlink{D_96}{\color{red}$\rm D\_96$}  & \hyperlink{D_97}{\color{red}$\rm D\_97$}  & {--} & {--} & \hyperlink{D_31}{\color{red}$\rm D\_31$}  & {--} & {--} & {--} & {--} & {--} & {--} & {--} & \hyperlink{D_28}{\color{blue}$\rm D\_28$} \\
\hline
$\rm D^{+}$ & \hyperlink{D_01}{\color{red}$\rm D\_01$}  & {--} & \hyperlink{D_04}{\color{red}$\rm D\_04$} \hyperlink{D_12}{\color{red}$\rm D\_12$}  & \hyperlink{D_33}{\color{red}$\rm D\_33$} \hyperlink{D_74}{\color{red}$\rm D\_74$}  & \hyperlink{D_05}{\color{red}$\rm D\_05$} \hyperlink{D_62}{\color{red}$\rm D\_62$} \hyperlink{D_63}{\color{red}$\rm D\_63$}  & {--} & {--} & \hyperlink{D_94}{\color{red}$\rm D\_94$}  & {--} & {--} & {--} & \hyperlink{D_36}{\color{red}$\rm D\_36$}  & \hyperlink{D_34}{\color{red}$\rm D\_34$} \hyperlink{D_75}{\color{red}$\rm D\_75$}  & {--} & \hyperlink{D_86}{\color{blue}$\rm D\_86$}  & {--} & \hyperlink{D_43}{\color{blue}$\rm D\_43$} \hyperlink{D_70}{\color{blue}$\rm D\_70$} \hyperlink{D_89}{\color{blue}$\rm D\_89$}  & {--} & {--} & {--}\\
\hline
$\rm D_2$ & \hyperlink{D_54}{\color{red}$\rm D\_54$} \hyperlink{D_71}{\color{red}$\rm D\_71$}  & \hyperlink{D_45}{\color{red}$\rm D\_45$} \hyperlink{D_46}{\color{red}$\rm D\_46$} \hyperlink{D_89}{\color{red}$\rm D\_89$}  & \hyperlink{D_48}{\color{red}$\rm D\_48$} \hyperlink{D_92}{\color{red}$\rm D\_92$}  & {--} & \hyperlink{D_93}{\color{red}$\rm D\_93$}  & {--} & {--} & \hyperlink{D_100}{\color{red}$\rm D\_100$}  & \hyperlink{D_98}{\color{red}$\rm D\_98$} \hyperlink{D_99}{\color{red}$\rm D\_99$}  & {--} & {--} & {--} & {--} & \hyperlink{D_90}{\color{red}$\rm D\_90$}  & {--} & {--} & {--} & {--} & {--} & {--}\\
\hline
$\rm D_2^{+}$ & {--} & {--} & \hyperlink{D_41}{\color{red}$\rm D\_41$} \hyperlink{D_42}{\color{red}$\rm D\_42$} \hyperlink{D_43}{\color{red}$\rm D\_43$}  & \hyperlink{D_82}{\color{red}$\rm D\_82$} \hyperlink{D_83}{\color{red}$\rm D\_83$}  & {--} & {--} & {--} & {--} & {--} & {--} & {--} & \hyperlink{D_86}{\color{red}$\rm D\_86$}  & \hyperlink{D_84}{\color{red}$\rm D\_84$} \hyperlink{D_85}{\color{red}$\rm D\_85$}  & {--} & {--} & {--} & {--} & {--} & {--} & \hyperlink{D_36}{\color{blue}$\rm D\_36$} \\
\hline
$\rm HD$ & \hyperlink{D_59}{\color{red}$\rm D\_59$} \hyperlink{D_60}{\color{red}$\rm D\_60$} \hyperlink{D_61}{\color{red}$\rm D\_61$}  & \hyperlink{D_06}{\color{red}$\rm D\_06$} \hyperlink{D_24}{\color{red}$\rm D\_24$} \hyperlink{D_64}{\color{red}$\rm D\_64$}  & \hyperlink{D_10}{\color{red}$\rm D\_10$} \hyperlink{D_65}{\color{red}$\rm D\_65$}  & {--} & \hyperlink{D_91}{\color{red}$\rm D\_91$}  & {--} & \hyperlink{D_55}{\color{red}$\rm D\_55$}  & \hyperlink{D_66}{\color{red}$\rm D\_66$}  & \hyperlink{D_67}{\color{red}$\rm D\_67$} \hyperlink{D_68}{\color{red}$\rm D\_68$} \hyperlink{D_69}{\color{red}$\rm D\_69$}  & {--} & {--} & \hyperlink{D_47}{\color{red}$\rm D\_47$}  & {--} & \hyperlink{D_44}{\color{red}$\rm D\_44$} \hyperlink{D_87}{\color{red}$\rm D\_87$} \hyperlink{D_88}{\color{red}$\rm D\_88$}  & {--} & {--} & {--} & {--} & {--} & \hyperlink{D_07}{\color{blue}$\rm D\_07$} \\
\hline
$\rm HD^{+}$ & \hyperlink{D_14}{\color{red}$\rm D\_14$}  & {--} & \hyperlink{D_09}{\color{red}$\rm D\_09$} \hyperlink{D_21}{\color{red}$\rm D\_21$} \hyperlink{D_25}{\color{red}$\rm D\_25$}  & \hyperlink{D_78}{\color{red}$\rm D\_78$} \hyperlink{D_79}{\color{red}$\rm D\_79$}  & \hyperlink{D_15}{\color{red}$\rm D\_15$} \hyperlink{D_56}{\color{red}$\rm D\_56$}  & {--} & {--} & {--} & {--} & {--} & {--} & \hyperlink{D_39}{\color{red}$\rm D\_39$} \hyperlink{D_40}{\color{red}$\rm D\_40$} \hyperlink{D_70}{\color{red}$\rm D\_70$}  & \hyperlink{D_80}{\color{red}$\rm D\_80$} \hyperlink{D_81}{\color{red}$\rm D\_81$}  & {--} & {--} & {--} & {--} & {--} & {--} & \hyperlink{D_11}{\color{blue}$\rm D\_11$} \hyperlink{D_12}{\color{blue}$\rm D\_12$} \\
\hline
$\rm H_2D^{+}$ & \hyperlink{D_17}{\color{red}$\rm D\_17$} \hyperlink{D_18}{\color{red}$\rm D\_18$} \hyperlink{D_19}{\color{red}$\rm D\_19$}  & {--} & \hyperlink{D_20}{\color{red}$\rm D\_20$}  & {--} & \hyperlink{D_57}{\color{red}$\rm D\_57$}  & {--} & {--} & {--} & {--} & {--} & {--} & {--} & {--} & {--} & {--} & {--} & {--} & {--} & {--} & {--}\\
\hline
$\rm \gamma$ & {--} & \hyperlink{He_18}{\color{orange}$\rm He\_18$}  & \hyperlink{H_02}{\color{red}$\rm H\_02$}  & \hyperlink{H_04}{\color{red}$\rm H\_04$}  & \hyperlink{H_14}{\color{red}$\rm H\_14$} \hyperlink{H_15}{\color{red}$\rm H\_15$}  & \hyperlink{H_08}{\color{red}$\rm H\_08$} \hyperlink{H_26}{\color{red}$\rm H\_26$}  & \hyperlink{H_21}{\color{red}$\rm H\_21$}  & \hyperlink{He_02}{\color{red}$\rm He\_02$} \hyperlink{He_18}{\color{orange}$\rm He\_18$}  & \hyperlink{He_07}{\color{red}$\rm He\_07$}  & {--} & \hyperlink{He_05}{\color{red}$\rm He\_05$} \hyperlink{He_14}{\color{red}$\rm He\_14$}  & \hyperlink{D_02}{\color{red}$\rm D\_02$}  & \hyperlink{D_53}{\color{red}$\rm D\_53$}  & {--} & \hyperlink{D_51}{\color{red}$\rm D\_51$} \hyperlink{D_52}{\color{red}$\rm D\_52$}  & \hyperlink{D_50}{\color{red}$\rm D\_50$}  & \hyperlink{D_49}{\color{red}$\rm D\_49$}  & \hyperlink{D_13}{\color{red}$\rm D\_13$} \hyperlink{D_26}{\color{red}$\rm D\_26$}  & {--} & {--}
\\
\hline
\end{tabular}}
\end{center}
{\tiny {\bf Note}: Red and orange labels indicate reactions where reactants are destroyed, while blue and cyan show reactions where reactants are synthesized. Red and blue correspond to reactions with four reactants, and other colours to those with more.}
\end{table}

\end{appendix}

\end{document}